# Extreme Optical Field Confinement and Enhancement in a Plasmonic Picopatch within a Nanoparticle-on-Mirror Resonator

Jinna He,[1,2] Mario Zapata-Herrera,[2,3] Xabier Arrieta,[2,4] Mingli Wan,[1] Yuan Zhang,[5,6] Javier Aizpurua, [3,4,7] and Ruben Esteban[*2,3]

[1]School of Electric and Mechanical Engineering, Pingdingshan University, Pingdingshan, 467000, China

[2]Centro de Física de Materiales (CFM-MPC) , CSIC-UPV/EHU, Paseo Manuel de Lardizabal 5, 20018 San Sebastián, Spain

[3]Donostia International Physics Center (DIPC), Paseo Manuel de Lardizabal 4, 20018 San Sebastián, Spain

[4]Dpt. of Electricity and Electronics, University of the Basque Country (UPV/EHU), 48940, Leioa

[5]Henan Key Laboratory of Diamond Materials and Devices, Key Laboratory of Material Physics, Ministry of Education, School of Physics, Zhengzhou University, Zhengzhou 450052, China

[6]Institute of Quantum Materials and Physics, Henan Academy of Sciences, Zhengzhou 450046, China

[7]Ikerbasque, Basque Foundation for Science, María Díaz de Haro 3, 48013 Bilbao, Spain

*ruben.esteban@ehu.eus

## Abstract

Plasmonic resonances in metallic nanogaps can confine light into nanometric regions, but reaching modes of volume $\approx 1\ nm^3$ remains challenging. We present a detailed theoretical analysis of the optical modes of an experimentally-motivated nanoresonator configuration that contains a picopatch formed by the lifting of a few gold atoms in the gap of a Nanoparticle-on-Mirror (NPoM) structure. Classical simulations indicate that the plasmonic modes associated with the picopatch geometry can confine light to extremely small regions and are highly sensitive to the picopatch size and shape, enabling broad tunability. Furthermore, these modes can couple strongly with other nanocavity modes of the structure, as identified from a clear anti-crossing behavior of the polaritonic resonances. Remarkably, up to ≈ 2000-fold electric field enhancement inside the picopatch and tiny effective mode volumes approaching ≈ 1 $nm^3$ are obtained. Further, changing the morphology of the picopatch does not modify the qualitative findings, and increasing the absorption losses in the classical simulations, to mimic quantum (non-local) effects in the metal permittivity, decreases the electric field enhancement only moderately. Compared to the standard picocavities formed by single-atom protrusions, picopatches are an intriguing alternative to reach extreme optical field confinement and enhancement in plasmonic cavities.

## 1. Introduction

Optical modes can confine light to small regions and are very attractive to miniaturize optical devices and simultaneously increase light-matter interaction. Notably, localized surface plasmons induced by the hybridization of optical fields and collective electronic excitations in metallic nanoparticles enable strong absorption and scattering of light, as well as light confinement well below the diffraction limit.[1,2] The plasmonic properties of these metallic nanoparticles can be altered by changing their geometry,[3] and thus a variety of metallic (plasmonic) nanostructures, from single nanoparticles of various geometries[4-7] to complex arrangements of multiple building blocks,[8-10] have been widely explored in recent years to achieve optimized field confinement and enhancement, for example.[11] In this context, a canonical configuration is the NanoParticle-on-Mirror (NPoM) optical resonator,[12-15] which can confine light very efficiently into the gap that separates a metallic nanoparticle from a metallic substrate (the mirror). An exquisitely defined NPoM gap thickness down to less than 1 nm,[16] typically set by a molecular monolayer or two-dimensional material that serve as a spacer, leads to several-hundred-fold electric field enhancements in the gap. The effective mode volume in the gaps of NPoM structures is tens to hundreds of cubic nanometers, about 6-7 orders of magnitude below the diffraction limit,[17] making the NPoM geometry a robust platform for a wide range of applications including Surface-Enhanced Raman or Fluorescence Spectroscopy (SERS and SEFS), nonlinear optics, plasmonic sensing, plasmon-exciton ultra-strong coupling, optoelectronic devices, optomechanics and quantum optics.[15,18]

Plasmonic resonances in the NPoM structures with ultrathin nanogaps are very sensitive to the precise structural details of the gap.[19-21] Notably, it has been demonstrated that atomic protrusions in these gaps can localize electromagnetic fields into regions ('hot spots') down to sub-nanometric dimensions ('picocavities').[22-27] Similar effects due to the presence of picocavities in a tip can be also clearly observed in scanning tunneling microscopy (STM) experiments that demonstrate spectral imaging with spatial resolution below 1 nm.[28-30] However, although the volume of the hot spot in these 'picocavities' is often smaller than 1 nm$^3$, the volume of the electromagnetic mode in a quantum-mechanical picture (see definition of this volume below) can be much larger, except for very specific configurations, such as those studied in Refs. [26, 31].

Picocavities are spontaneously formed in NPoM resonators by random motions of single (or very few) surface metallic atoms under intense laser illuminations, as sketched in Fig. 1(a). These motions can be induced by the thermal energy or optical forces. On the other hand, ultra-high-vacuum transmission electron microscopy (TEM) experiments[32] have also observed the collective delamination of several metallic atoms (e.g., patches of 5 ~ 10 atoms wide), which forms a tiny (subatomic thick and nanometer wide) empty slot between them and the metallic surface (Fig. 1(b)). Here we call these structures 'picopatches', even if their lateral dimensions can be slightly larger than 1 nm, in analogy to the picocavities formed by isolated atomic protrusions. Our previous work[33] suggested that the formation of picopatches in the NPoM nanoresonators (Fig. 1(c)) can be behind experimental observations of transient and spectrally broad emission (flares) from plasmonic nanostructures,[34-36] due to strong localization and enhancement of the fields. However, this previous work was focused on the analysis of the laterally infinite metal-insulator-metal-insulator (MIMI) planar waveguide systems, with only preliminary simulations of a full 3-dimensional NPoM resonator with a picopatch.

In this work, we show the potential of a NPoM containing a picopatch as an extreme field confiner in realistic configurations (Fig. 1(c)), through a detailed theoretical analysis of their optical properties. We first discuss the existence of the plasmonic modes associated with the picopatch and characterize their properties based on calculations of a planar MIMI configuration. We then analyze the optical response of the full system depicted in Fig. 1(c), focusing on the extreme field enhancement and confinement that can be reached. We also analyze the tunability of the modes in this structure with modifications of the structural parameters as well as the emergence of hybrid modes due to the strong coupling between the modes confined in the picopatch and other modes confined in the NPoM nanocavity. The coupling strength that characterizes this interaction is extracted from a coupled harmonic oscillator model, and the mode volume of the hybrid modes is evaluated to quantify the extreme, close to 1 $nm^3$, field confinement. Last, we examine the robustness of the results by considering modifications of the exact picopatch configuration and of the absorption losses. These modifications are motivated by the impossibility of knowing or controlling the precise experimental geometry, and by the fact that the local classical permittivity used in the simulations becomes less well justified at tiny scales due to quantum (non-local) effects.

## 2. Morphology

The configuration of the NPoM resonator containing the picopatch is shown schematically in Fig. 1(c), with the zoom on the right focusing on the structural details of the picopatch itself. A bare NPoM resonator (corresponding to the same geometry but without the picopatch) is used as a reference and consists of a truncated Au spherical nanoparticle of radius $r_{\mathrm{sphere}}$ = 40 nm with a flat bottom facet of radius $r_{facet}$ = 30 nm separated from a semi-infinite Au substrate underneath by a dielectric spacer of thickness $d$ = 1.1 nm and permittivity $\varepsilon_{\mathrm{d1}}$ = 2.1 (all values given in this section are those used in the calculations except when explicitly mentioned otherwise). The spacer covers the whole Au substrate and can represent a molecular monolayer, for example. Inside the NPoM gap, the picopatch is formed by a few Au atoms that are lifted from the bottom metallic substrate, so that they are separated $\delta$ = 0.2 nm from it. These lifted atoms are placed below the center of the flat facet and present a circular cross section in the horizontal x-y plane parallel to the Au substrate. The thickness of the lifted Au atoms is set to be $t$ = 0.235 nm and corresponds to the thickness of an Au monolayer (as set by the Au (111) layer spacing). The slot between the Au monolayer and the underlying bulk Au is filled up by vacuum with permittivity $\varepsilon_{\mathrm{d2}}$ = 1.0. The slot has a typical radius $r_{\mathrm{slot}}$ = 1.435 nm, corresponding to a lateral width of ~ 10 Au atoms. The picopatch corners are rounded to avoid unphysical divergences in the classical calculations of field enhancements. The whole structure is rotationally symmetric and is illuminated by a plane wave incoming with an incidence angle of $45^0$ with respect to the normal of the substrate (z-axis) and polarized in the x-z plane of incidence (p-polarization) (see the axis of coordinates, illumination and details of the geometry in the sketches in Fig. 1(c)).

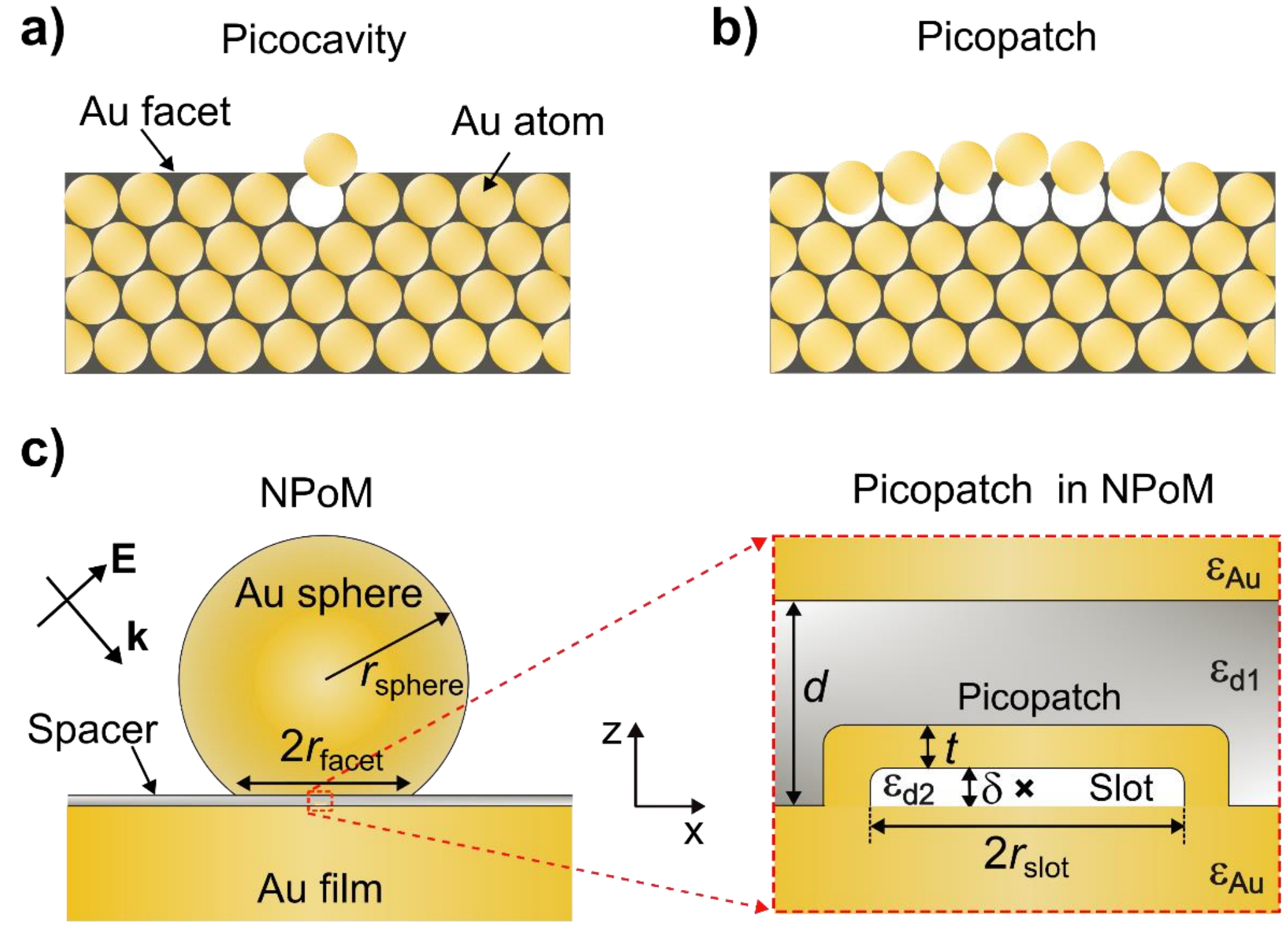

**Fig. 1 NPoM resonator with picopatch.** (a, b) Schematics of an Au interface, showing (a) a single adatom protruding from the surface (picocavity) and (b) delamination of a few atoms in the top Au monolayer (picopatch). White circles represent the original position of the gold atoms before they move to form the picocavity or picopatch. (c) Sketch of the NPoM resonator with a picopatch considered in this work, with the zoom on the right focusing on the structural details of the picopatch itself. The NPoM consists of a truncated Au nanosphere of radius $r_{sphere}$ = 40 nm over a semi-infinite Au substrate, with a $d$ = 1.1 nm thick dielectric spacer of permittivity $\varepsilon_{d1}$ = 2.1 between them forming a nanogap that contains the picopatch. The truncated Au sphere presents a circular flat facet at the bottom of radius $r_{facet}$ = 30 nm, which corresponds to the upper interface of the nanogap. The picopatch is situated in the nanogap, below the center of the flat facet, and it consists of a small circular section of an Au monolayer (thickness $t$ = 0.235 nm and radius $r_{slot}$), at a $\delta$ = 0.2 nm distance from the underlying substrate, leaving a vacuum slot in the middle (permittivity $\varepsilon_{d2}$ = 1.0). A vertical wall of the same thickness $t$ connects the edges of the monolayer and the substrate (see zoom in the right panel for details). The full geometry is rotationally symmetric around the vertical z-direction, and the illumination is a p-polarized wave incoming in the x-z plane at a $45^0$ angle with respect to the vertical z direction (x and z directions are indicated in (c), with the nanogap center corresponding to x = y = 0 and z = 0.55 nm). The black cross at the center of the slot (x = y = 0 and z = 0.1 nm) indicates the position where the electric fields are often evaluated and where a dipole is placed when calculating the Purcell factor in the main text.

## 3. Methodology

The classical optical response of the NPoM resonator that contains the picopatch is obtained using full-wave simulations implemented in COMSOL Multiphysics®.[37] The computational model contains the structure under consideration, and the sphere-shaped perfectly matched layers (PML) are applied in all directions to eliminate nonphysical boundary reflections. Since the structure involves a semi-infinite gold substrate, an analytical background field instead of a plane wave is used as an excitation.[38] This background field corresponds to that excited by a plane wave in a layered system formed by the gold substrate, the $d$ = 1.1 nm thick dielectric layer and the vacuum on top. The permittivity of gold $\varepsilon_{Au}$ is taken from experimental data,[39] except when stated otherwise. The scattering cross section is obtained by computing the surface integral of the scattered Poynting vector in the upper semi-sphere at the internal boundary of the PML layer. The absorption cross section is obtained by calculating the volume integral of the resistive losses in the metallic regions occupied by the gold nanosphere and the lifted gold monolayer. Note that we do not extend this integral to the metallic substrate, so that the absorption obtained is smaller than if the substrate were included. This simplification should not affect the spectral shape and thus the identification and analysis of the modes. Considering the extremely small dimensions of the picopatch, the convergence of the calculations was carefully verified by varying the size of the whole simulated domain, the mesh grid and the thickness of the PML layer (see further simulation details and convergence tests in section S2 of the Supplementary Information).

## 4. Results and Discussion

### 4.1 MIM and MIMI planar waveguides

To relate the complex structure of plasmonic modes in the picopatch structure to simpler plasmonic waveguiding modes, we first consider the plasmonic modes in laterally-infinite Metal-Insulator-Metal (MIM) and Metal-Insulator-Metal-Insulator (MIMI) planar waveguides, as described by classical electromagnetic calculations within local dielectric response theory. The justification of this simple classical treatment is discussed at the end of this section. Crucially, the bare NPoM nanocavity (without the picopatch) supports Fabry-Pérot-like transverse cavity plasmons (TCPs) along the 1.1 nm-thick gap that originate from the reflection at the gap edges of the surface plasmon propagating in it.[40-45] The properties of this propagating surface plasmon can be obtained by studying an infinite MIM configuration that matches the local geometry of the NPoM gap, where the 'insulator' in the 'Metal-Insulator-Metal' structure is the dielectric in the gap. In a similar manner, the picopatch forms locally a MIMI (gold/vacuum slot/gold monolayer/dielectric) multilayer structure that supports propagating plasmons, which are reflected at the picopatch edges and lead to an additional set of Fabry-Pérot-like TCP modes. We do not include an additional metallic layer in the MIMI waveguide, corresponding to the nanoparticle at the top, because its effect on the dispersion is relatively weak. Understanding the dispersion of the MIM and MIMI modes is crucial to later analyze the optical response of the full system (that is, the NPoM containing a picopatch).

The infinite MIM waveguide corresponding to the local configuration of the nanogap in the NPoM nanocavity is depicted in the top inset of Fig. 2. This configuration is composed of a gold semi-infinite layer (permittivity $\varepsilon_{Au}$), a dielectric layer (permittivity $\varepsilon_{d1}$ = 2.1) of typical thickness $d$ = 1.1 nm, and a second gold semi-infinite layer. Using classical electromagnetic theory, the dispersion of these MIM waveguide modes[46] can be obtained by solving the following equation,

$$\tanh\left(\frac{id}{2}\sqrt{{k_0}^2\varepsilon_{d1} - q^2}\right) = \frac{\varepsilon_{d1}}{\varepsilon_{Au}}\sqrt{\frac{q^2 - \varepsilon_{Au}{k_0}^2}{q^2 - \varepsilon_{d1}{k_0}^2}}\ , \tag{1}$$

where $k_0$ is the vacuum wavevector, and $q$ is the in-plane plasmonic wavevector of the propagating MIM plasmonic mode, i.e., the component parallel to the layers (x-axis) of the wavevector of the plasmonic excitation (see section S1.1 of the Supplementary Information for the derivation of Eq. (1)).

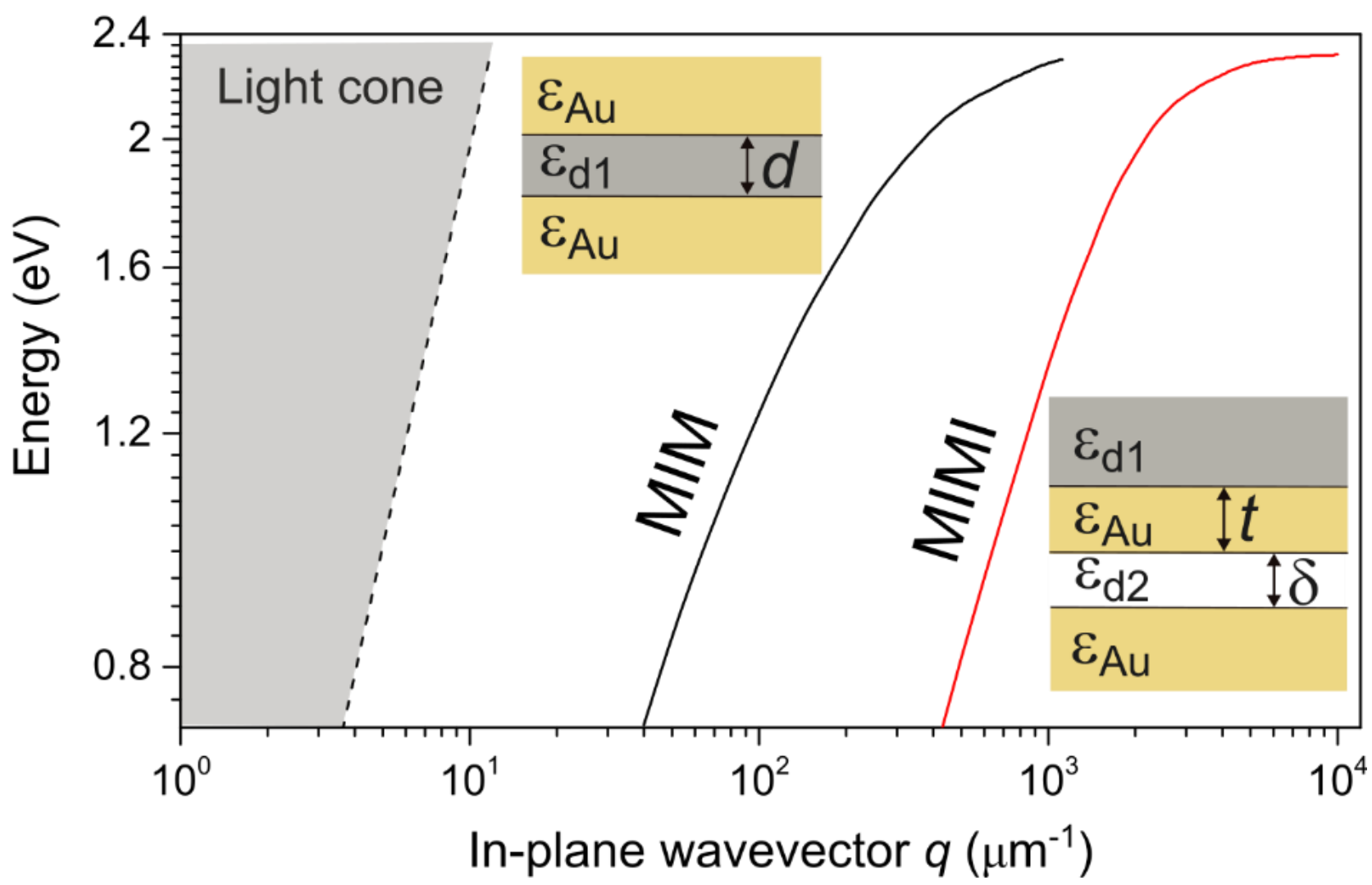

**Fig. 2 Laterally infinite MIM and MIMI waveguides.** Dispersion relationships of propagating plasmon modes in the planar MIM (black line) and MIMI (red line) configurations shown in the insets. The light cone of free photons in vacuum is represented by the gray area. Notice the logarithmic scale used here. The MIM structure consists of two semi-infinitely-thick Au films separated by a dielectric layer of permittivity $\varepsilon_{d1}$ = 2.1 and thickness $d$ = 1.1 nm (top inset). The MIMI structure consists of a semi-infinitely-thick Au film, a vacuum layer of permittivity $\varepsilon_{d2}$ = 1.0 and thickness $\delta$ = 0.2 nm, a thin gold layer of thickness $t$ = 0.235 nm, and a semi-infinitely-thick dielectric layer of permittivity $\varepsilon_{d1}$ = 2.1 (bottom-right inset). The MIM and MIMI dispersions are plotted as a function of the in-plane wavevector $q$. The dispersion of the MIM and MIMI modes tends towards zero energy for vanishingly small wavevectors (not evident in the figure because of the logarithmic scale).

On the other hand, the MIMI multilayer, corresponding to the local structure in the picopatch (bottom-right inset in Fig. 2), consists of a thin gold layer of thickness $t$ = 0.235 nm separated by a $\delta$ = 0.2 nm thick vacuum layer (permittivity $\varepsilon_{d2}$ = 1.0) from a semi-infinite Au surface (permittivity $\varepsilon_{Au}$) underneath. On top of the thin Au layer there is a semi-infinite dielectric layer (permittivity $\varepsilon_{d1}$ = 2.1). The classical non-retarded dispersion of the MIMI plasmons confined in the subatomic-thick slot layer can be obtained by considering that the in-plane wavevector $q$ is very large[33] (see section S1.2 of the Supplementary Information). We obtain that the plasmon MIMI modes verify,

$$\left(\frac{\varepsilon_{Au}-\varepsilon_{d1}}{\varepsilon_{Au}+\varepsilon_{d1}}\right)\left(\frac{\varepsilon_{Au}-\varepsilon_{d2}}{\varepsilon_{Au}+\varepsilon_{d2}}\right)\left[\left(\frac{\varepsilon_{Au}+\varepsilon_{d1}}{\varepsilon_{Au}-\varepsilon_{d1}}\right)\left(\frac{\varepsilon_{Au}-\varepsilon_{d2}}{\varepsilon_{Au}+\varepsilon_{d2}}\right)\right.$$
$$\left.-\left(\frac{(\varepsilon_{Au}+\varepsilon_{d1})(\varepsilon_{Au}-\varepsilon_{d2})}{(\varepsilon_{Au}-\varepsilon_{d1})(\varepsilon_{Au}+\varepsilon_{d2})}-e^{-2qt}\right)\left(1-e^{-2q\delta}\right)\right]-1=0, \quad (2)$$

with $q$ the in-plane plasmonic wavevector of the propagating MIMI plasmonic mode.

The dispersion curves of the MIM and MIMI (i.e., the energy of the corresponding propagating surface plasmons as a function of the in-plane wavevector $q$) are obtained from Eq. (1) and Eq. (2) and are plotted in Fig. 2 as black and red lines, respectively. For reference, the light cone of the free space photons ($\omega \geq ck_0$ with $k_0$ wavevector in vacuum) is represented by the gray area. The in-plane wavevector of surface plasmons propagating in the 1.1 nm-thick MIM waveguide is one order of magnitude larger than the vacuum wavevector for the energies considered,[47] $q \approx 10k_0$. Remarkably, the MIMI in-plane wavevector is an additional order of magnitude larger than that of the MIM, and thus presents a hundred-fold wavevector mismatch when compared to that of the free photons, $q \approx 100k_0$. Such a huge $q$ is behind the capability of picopatches to confine light into extremely small regions. However, the large mismatch in $q$ between the MIMI mode and free photons also indicates that this plasmonic mode cannot be excited directly by incident light, making it challenging to optically exploit this mode in a planar system. A nanoresonator locally containing a MIMI can be used to overcome this limitation, as shown in the next section where we analyze the optical response of a localized MIMI structure (the picopatch) located within a large NPoM resonator. We note further that the MIMI plasmon dispersion exhibits a nearly linear dependence on the in-plane wavevector $q$ up to energy ~ 2.2 eV (red line), indicating that this mode can be categorized as an acoustic excitation.[48-50]

This analysis has considered a local classical permittivity of gold. The validity of this treatment for the very small thickness and separation distances involved in the MIMI structure was examined using full quantum ab-initio calculations in our recent work.[33] The dispersion obtained from the quantum theory was found to match the classical dispersion surprisingly well for thicknesses larger than a very small threshold separation distance, $\delta \gtrsim 0.2$-$0.3$ nm, and the field spatial distribution also showed good agreement for the thicknesses larger than this value. These results provided compelling evidence that the adopted classical electromagnetic theory is capable of reproducing well many of the features of the ab-initio calculations, even for an extremely thin gap as far as it is larger than the $\delta \approx$ 0.2-0.3 nm threshold. This justifies the use of full-wave classical electromagnetic simulations in the following. On the other hand, the spectral width of the peak in the ab-initio calculations was broader (sometimes narrower) than that of the classical results, depending on the in-plane wavevector $q$, which indicates changes in the absorption losses that will affect the field enhancement in the nanoresonators. Notably, the local classical description does not include electron tunneling and non-local quantum effects such as Landau damping and electron spill-out that can increase losses,[51-54] and thus be responsible for the modifications in plasmon widths. The effect of the possible underestimation of the absorption losses in the classical calculations is considered at the end of this work.

**4.2. Plasmonic properties of bare NPoM and NPoM with picopatch**

To better identify the effect of the picopatch, we first consider the optical response of bare NPoM resonators, corresponding to the structure in Fig. 1(c) after removing the picopatch. As previously introduced, the system is illuminated by a p-polarized plane wave incoming at a $45^0$ angle. The simulated spectra of the scattering cross section, the absorption cross section and the near-field enhancement $|\boldsymbol{E}/\boldsymbol{E_0}|$ at the nanogap center are shown by the red dashed lines in Figs. 3(a-c) ($|\boldsymbol{E}|$ amplitude of the induced electric field; $|\boldsymbol{E_0}|$ corresponding value for the incident plane wave that excites the system). Following the nomenclature adopted previously,[19,20] we attribute the peak at the wavelength $\lambda \approx 660$ nm to the fundamental longitudinal antenna plasmon (LAP) mode ($l_1$) and the remaining peaks to transverse cavity plasmons (TCPs) of different orders ($s_{mn}$). The former ($l_1$) is characterized by the excitation of currents normal to the substrate ($z$ direction). The latter ($s_{mn}$) are Fabry-Pérot-like modes induced by the reflection at the gap edges of the plasmons confined to the gap and propagating parallel to the gap interfaces (x-y plane), with the wavevector given by the MIM dispersion (black line in Fig. 2). The $s_{mn}$ mode is characterized by a spatial electric field distribution in the gap that approximately follows $E_z(\rho,\varphi) \approx J_m(a'_{mn}\,\rho/r_{facet})\cos(m\varphi)$, with $E_z$ the z component of the field, $J_m$ the Bessel function of the first kind, $a'_{mn}$ the n$^{th}$ root of the derivative of this function, and $\rho$ and $\varphi$ the radial and azimuthal coordinates[45,55] (a related but different nomenclature is used in, e.g., Refs. [20, 56]). Thus, the subindexes (m, n) in $s_{mn}$ refer to the symmetry of the different orders of the modes, with m the number of nodes in the azimuthal direction (for $\varphi$ in the range [0, π]) and n the number of nodes in the radial direction ($\rho$ in the range [0, $r_{facet}$]). The assignment of these eigenmodes of the bare NPoM nanocavity is discussed in more detail below and in section S3 of the Supplementary Information (Figs. S4 and S5).

The signature of the $s_{mn}$ modes can be identified more clearly in the absorption cross section (Fig. 3(b), red dashed line), with the $s_{02}$, $s_{12}$, $s_{01}$ and $s_{11}$ modes appearing as peaks at increasing wavelengths. In contrast, the $s_{11}$ and $s_{12}$ modes are not distinguishable in the scattering spectra (Fig. 3(a), red dashed line) because these modes radiate weakly ('dark modes'), nor in the near-field enhancement evaluated in the gap center (Fig. 3(c), red dashed line) because these modes present a zero value at this position due to their symmetry. Additionally, the $s_{01}$ mode appears in the scattering as a Fano-like resonance due to the interference with the background response coming from higher-energy modes.[57-59] A similar effect could also explain why the $s_{02}$ peak appears at a slightly different wavelength in the scattering ($\lambda \approx 725$ nm) and in the absorption ($\lambda \approx 710$ nm) spectra. We further highlight that the maximum electric field enhancement for bare NPoM that appears at the peak associated with the $s_{02}$ mode is $|\boldsymbol{E}/\boldsymbol{E_0}| \approx 200$.

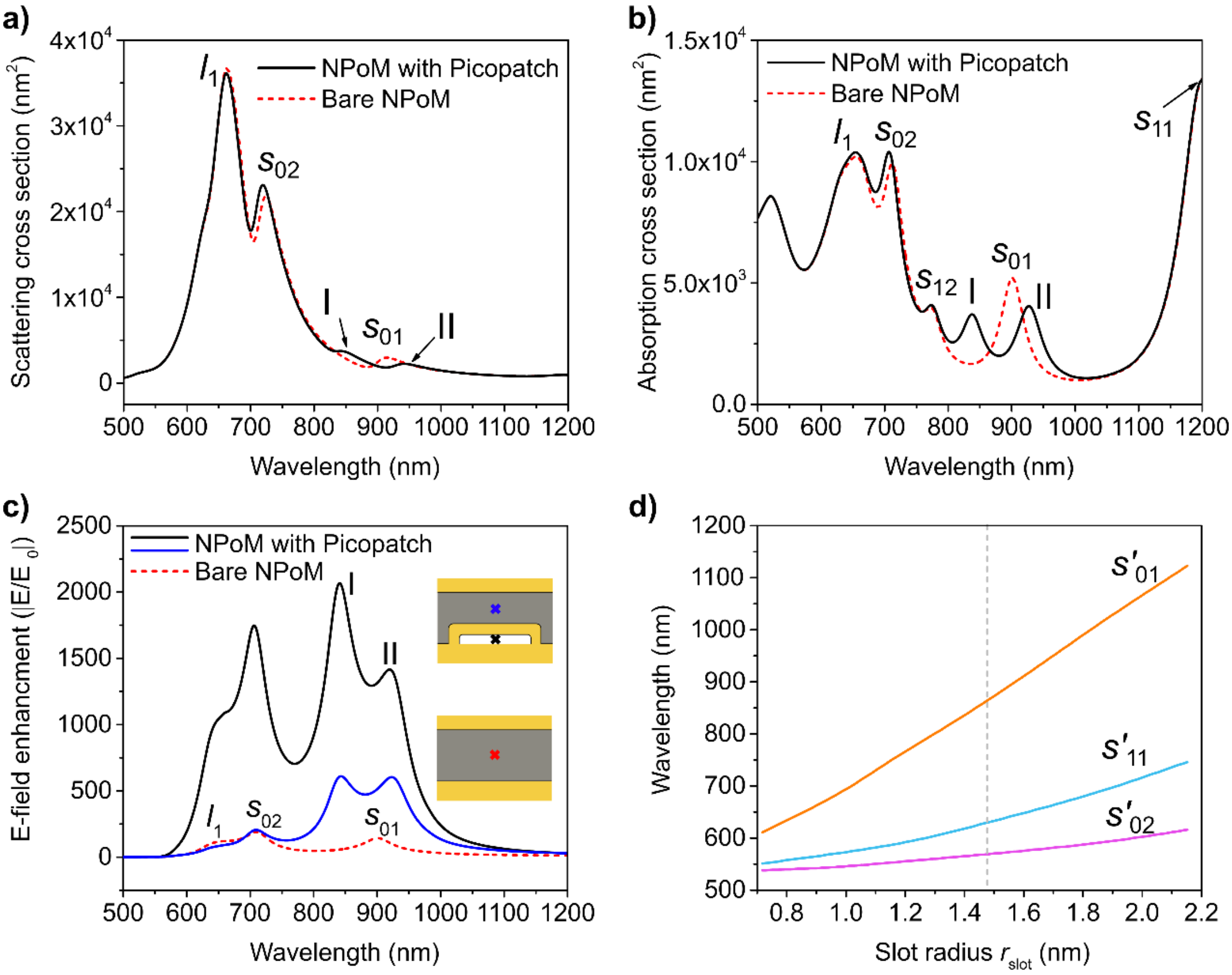

**Fig. 3 Optical response of bare NPoM and NPoM containing a picopatch**. (a) Scattering and (b) absorption cross section spectra (in nm$^2$), and (c) E-field enhancement spectra ($|E/E_0|$) in the absence (bare NPoM: red dashed line) and presence (NPoM with picopatch; black and blue solid lines) of the picopatch. The field enhancement is obtained at the center of the gap of the bare NPoM nanocavity without picopatch (red dashed line; position marked by red cross in the bottom inset) or, when the picopatch is included, at the center of the slot formed between the metal substrate and the thin metal layer of the picopatch (solid black line; black cross in the top inset and in Fig. 1(c)) and near the center of the nanogap (blue line, blue cross in the top inset; see Fig. S10 in the Supplementary information for discussion of the exact position of the evaluated fields). Results in (a-c) are obtained for the following parameters: $r_{sphere}$ = 40 nm, $r_{facet}$ = 30 nm, $d$ = 1.1 nm, $\delta$ = 0.2 nm, $t$ = 0.235 nm, and $r_{slot}$ = 1.435 nm. (d) Resonant wavelength of the modes of the picopatch ($s'_{mn}$, picopatch modes) as a function of the slot radius $r_{slot}$ in the picopatch. The wavelength is obtained from analytical Eqs. (2) and (4), which do not include coupling with other modes. The values of the parameters other than $r_{slot}$ are the same as in (a-c). Vertical dashed line corresponds to the value of $r_{slot}$ = 1.435 nm used in (a-c).

We consider next the response of the NPoM containing a picopatch (sketched in Fig. 1(c)) under the same plane wave illumination. The spectra of the scattering and absorption cross sections, as well as the field enhancement spectra evaluated in the center of the picopatch slot (position marked by a black cross in the inset top and in Fig. 1(c)), are shown by black solid lines in Figs. 3(a-c). The scattering (Fig. 3(a)) and absorption (Fig. 3(b)) cross section spectra of the NPoM containing a picopatch are generally very similar to those of the bare NPoM (compare black lines and red-dashed lines). However, in the spectral region near $\lambda$ = 900 nm, the $s_{01}$ peak in the bare NPoM spectra splits into two (marked as "I" and "II") when the picopatch is included. This splitting is due to the strong coupling between the $s_{01}$ nanocavity mode and a new Fabry-Pérot-like TCP mode associated with the picopatch ($s'_{mn}$ picopatch modes in the following), resulting in these two hybrid modes. It is remarkable that this substantial change, which may be accessible in far-field experiments and thus be used to detect the formation of picocavities [34], can be induced by such a small change in the local geometry (extending over a height $t + \delta$ = 0.435 nm and a radius $r_{slot}$ + 0.235 nm = 1.67 nm, and thus a volume of only ≈ 3.8 nm$^3$). Additionally, the presence of the picopatch results in a small shift of the $s_{02}$ nanocavity mode.

The effect of the picopatch is much more dramatic when considering the local fields (Fig. 3(c)). The $s_{01}$ peak in the bare NPoM spectra evaluated in the center of the nanogap (red dashed line) near $\lambda$ = 900 nm splits again into two in the spectra calculated in the center of the slot when the picopatch is present (black solid line). Importantly, the electric field enhancement inside the slot is greatly enhanced by the picopatch, approximately one order of magnitude, up to $|E/E_0| \approx 2000$. Furthermore, the blue line in Fig. 3(c) shows that the picopatch also strongly enhances the field spectra, up to $|E/E_0| \approx 600$, at the region of the nanogap over the picopatch (position indicated by the blue cross in the top inset), which could be exploited in spectroscopy measurements of molecules placed in this region.

We analyze next in more detail the Fabry-Pérot-like $s_{mn}$ and $s'_{mn}$ transverse cavity plasmon (TCP) modes. As introduced above, both the NPoM nanogap and the subatomic-thick slot can behave as waveguides with finite boundaries, which thus discretize the continuum of propagating MIM or MIMI modes into localized plasmon states. These states emerge when the modes that propagate parallel to the nanocavity gap or the picopatch slot reflect at the gap or slot edges, respectively, and interfere constructively, analogously to the resonances of a Fabry-Perot resonator.[19,42,60,61] The resulting Fabry-Perot-like resonances contribute to two sets of TCP modes, associated with either the nanocavity ($s_{mn}$) or the picopatch ($s'_{mn}$). Both sets of modes are present in the NPoM resonator containing the picopatch. Ignoring the interaction between different modes, the resonant plasmonic wavelength $\lambda_{pl}$ of the set of nanocavity TCP modes ($s_{mn}$) can be obtained from Refs. [45] and [55], giving

$$\lambda_{pl} = \frac{2\pi r_{facet}}{a'_{mn}}, \tag{3}$$

where $r_{facet}$ is the radius of the bottom flat facet of the gold nanosphere, and $a'_{mn}$ is the n$^{th}$ root of the derivative of the Bessel function of order m, $J_m$. On the other hand, we approximate the resonant plasmonic wavelength $\lambda'_{pl}$ of the set of the picopatch TCP modes ($s'_{mn}$) as

$$\lambda'_{pl} = \frac{2\pi r_{slot}}{a_{mn} - \beta}. \tag{4}$$

Here, $r_{slot}$ is the radius of the subatomic-thick slot, and we consider the n$^{th}$ root of the Bessel function of order m ($a_{mn}$), instead of its derivative, and include a reflection phase[62-63] $\beta = \pi/4$. The difference between Eq. (3) and Eq. (4) can be understood from the contrast between the spatial field distribution of the two types of modes. The fields of the $s_{mn}$ modes present approximately a local maximum at the nanocavity edge, while the field of the $s'_{mn}$ modes is closer to a minimum at the edge of the slot (see Fig. 5(b) and 5(d) below). We attribute this difference to the contrast in the terminations of the cavity: the picopatch slot is ‘close’, i.e. with metals at the edges, contrary to the ‘open’ nanogap limited laterally by vacuum. We do not attempt to assess the validity of Eq. (4) for general systems, but we have found it suitable to model the response of the structure considered in this work. The resonant vacuum wavelength of the different modes can then be obtained theoretically using the dispersion of the waveguide modes (Eq. (1) or Eq. (2)) together with the resonant conditions (Eq. (3) or Eq. (4)).

Figure 3(d) shows the expected dependence of the resonant vacuum wavelength of the picopatch TCP modes on slot radius $r_{slot}$ for different orders (m, n), as given by the analytical Eq. (2) and Eq. (4) (the nanocavity modes are discussed in section S3 of the Supplementary Information). The picopatch modes ($s'_{mn}$) redshift clearly with increasing $r_{\text{slot}}$, especially for the lowest-order $s'_{01}$ mode. When $r_{\text{slot}}$ is increased from ~ 0.7 nm to ~ 2.1 nm (a very small change in geometry), the $s'_{01}$ mode can be tuned from a vacuum wavelength of $\lambda \approx 610$ nm to $\lambda \approx 1100$ nm, covering wavelengths from the visible to well into the near-IR region. The predicted presence of plasmonic modes for such extremely small lateral size of the picopatch ($r_{\text{slot}} \approx$ 1-2 nm) is remarkable, and is only possible because of the very small thickness of the picopatch.

For the slot radius $r_{\text{slot}} = 1.435$ nm used in Figs. 3(a-c), the predicted $s'_{01}$ mode is found near $\lambda = 850$ nm in Fig. 3(d) (as marked by the vertical dashed line). This wavelength is close to the $s_{01}$ mode of the NPoM nanocavity. Thus, the peaks observed in the spectra of Figs. 3(b, c) at $\lambda \approx$ 830-930 nm (black solid lines) can be attributed to the coupling between the $s'_{01}$ and $s_{01}$ modes, which results in two new hybridized modes. Next, we discuss further this hybridization and the validity of the analytical model that leads to Eq. (2) and Eq. (4).

**4.3. Hybridization of nanocavity and picopatch modes**

We can exploit the tunability of the $s'_{01}$ picopatch TCP mode with changes of the picopatch size (specifically the slot radius $r_{\text{slot}}$). This approach serves to analyze in more detail the hybridization of this mode with other modes of the NPoM resonator. The 2D color maps in Figs. 4(a, b) show the dependence of the simulated absorption cross section and near-field enhancement ($|\boldsymbol{E}/\boldsymbol{E_0}|$) on the excitation wavelength $\lambda$ and slot radius $r_{\text{slot}}$ (the corresponding dependence of the scattering cross section spectra is given in Fig. S6 in the Supplementary Information). The spectra to the left of the color maps correspond to the reference results for the same system but without the picopatch (bare NPoM nanocavity). Horizontal dashed lines are also included in the 2D color maps to mark the position of the bare NPoM nanocavity modes ($l_1$ and $s_{mn}$), as extracted from the simulations in the absence of picopatch. Last, the theoretical wavelength of the picopatch mode ($s'_{01}$) as a function of $r_{slot}$, according to Eq. (2) and Eq. (4), is indicated by the diagonal dashed line.

For most values of $r_{slot}$, the NPoM nanocavity $s_{mn}$ modes are unaffected or weakly affected by the presence of the picopatch, as indicated by the presence of broad horizontal bands in the color maps. The observed modification and shift of these modes is a remarkable effect considering the tiny size of the picopatch, but the change is often small. An exception occurs for illumination wavelengths at $\lambda \approx$ 800-1000 nm and slot radius $r_{slot} \approx$ 1.4-1.8 nm. In this region, the spectral peaks show a clear anti-crossing behavior that can be understood as being due to the coupling between the $s_{01}$ nanocavity mode (horizontal dashed line) and the $s'_{01}$ picopatch mode (diagonal dashed line). We emphasize that the horizontal and diagonal dashed lines mark the evolution of these modes in the absence of coupling, obtained respectively from the simulation of the bare NPoM nanocavity without picopatch and from the theoretical model.

The anti-crossing is observed when the $s'_{01}$ mode is tuned to the wavelength of the $s_{01}$ mode, and is a consequence of the strong coupling between them,[64-69] giving rise to the two new hybrid modes labeled 'I' and 'II', as indicated in Figs. 3(a-c). Additionally, this hybridization leads to a particularly efficient coupling of the incoming light into the subatomic-thick vacuum slot in the picopatch, as the top Au nanoparticle acts as a nanoantenna that feeds light into the NPoM nanogap mode coupled to the picopatch. This behavior results in the strongly enhanced E-fields shown in Fig. 4(b). Very strong electric-field enhancements are also found when the $s'_{01}$ mode is tuned close to the wavelength of the $s_{02}$ nanocavity mode, but in this case the presence of an anti-crossing behavior is less clear. In addition, the absorption map in Fig. 4(a) exhibits a crossing behavior between the $s'_{01}$ picopatch mode and the $s_{12}$ nanocavity mode around the wavelength of $\lambda \approx$ 750 nm and slot radius of $r_{slot} \approx$ 1.0 nm. This behavior is a signature of the absence of (or weak) coupling between these modes, a consequence of the different symmetry of their associated near fields (compare the near fields in Fig. S5(b) and Fig. S7(a) in Supplementary Information).

To confirm the strong coupling between the $s'_{01}$ and $s_{01}$ modes, we fit the simulated absorption cross section spectrum of the NPoM with picopatch in the anti-crossing spectral region (near $r_{slot} \approx 1.5$ nm and $\lambda \approx 875$ nm) to the analytical expression derived from a coupled harmonic oscillator model,[70,71]

$$\sigma_{ext}(\omega) \propto \omega\, Im\left\{\frac{{\omega_s}^2 - \omega^2 - i\omega\gamma_s}{({\omega_g}^2 - \omega^2 - i\omega\gamma_g)({\omega_s}^2 - \omega^2 - i\omega\gamma_s) - 4g^2\omega^2}\right\}, \quad (5)$$

where Im{} refers to the imaginary part, $\omega_g$ and $\omega_s$ are the resonant frequencies of the $s_{01}$ and $s'_{01}$ modes, $\gamma_g$ and $\gamma_s$ are their corresponding damping rates, and $g$ is the coupling strength between them. These resonant frequencies and damping rates are those of the bare modes, i.e., the modes that would be present if the coupling could be switched off. In the fitting procedure, $\omega_g$ (corresponding to the $s_{01}$ mode in the nanogap) and $\gamma_g$ (corresponding to the damping rate of the $s_{01}$ mode) are allowed only small variations around the value obtained for the bare NPoM nanocavity ($\hbar\omega_g = 1.4$ eV and $\hbar\gamma_g = 0.035$ eV). On the other hand, $\omega_s$ (corresponding to the $s'_{01}$ mode in the picopatch slot), the coupling strength $g$, and the damping rate of $\gamma_s$ are fully free fitting parameters that vary with $r_{slot}$. Notice that two slightly different versions of the coupled oscillator model have been used in the literature,[72] but the results obtained for our system should not depend significantly on this choice. Additionally, Eq. (5) corresponds to the expression of the extinction spectra instead of the absorption, but this choice is not expected to result in any substantial difference in the values of coupling and anti-crossings extracted.

Figure 4(c) shows the fittings for $r_{slot} = 1.363$ - $1.865$ nm and wavelength between $\lambda =$ 800 - 1050 nm, with the simulated absorption spectra shown by the dots, and the result of the fit shown by solid lines. The fit is very satisfactory. The coupling strength $g$, resonant frequencies ($\omega_g$ and $\omega_s$), and both a half $(\gamma_g+\gamma_s)/2$ and a quarter $(\gamma_g+\gamma_s)/4$ of the total damping rates, extracted from all fittings, are plotted in Figs. 4(d, e). The black square dots in Fig. 4(d) show the small variation of the extracted resonant energy $\hbar\omega_g$ with respect to the resonant position of the $s_{01}$ nanocavity mode ($\hbar\omega_g$ = 1.4 eV, $\hbar$ reduced Planck's constant). In contrast, the extracted resonant energy $\hbar\omega_s$ (red circle dots in Fig. 4(d)) presents a nearly linear shift with the change of $r_{slot}$, in good overall agreement with the $s'_{01}$ picopatch mode energies obtained with the analytical model (Eq. (2) and Eq. (4)), which are also shown in Fig. 4(d) (blue solid line). This agreement further supports the validity of the analytical model to describe the (uncoupled) picopatch modes.

To determine whether we are in the weak or strong coupling regime, we compare in Fig. 4(e) the value of the extracted coupling strength $g$ with the damping rates of the two resonant modes. The coupling strength $g$ decreases moderately with increasing slot radius $r_{slot}$ (black square dots in Fig. 4(e)), but, more importantly, it is always very large, $\hbar g \approx$ 50-80 meV. Thus, the frequently-used criteria to identify the strong coupling regime, $g \geq (\gamma_g+\gamma_s)/4$, is always amply satisfied ($(\gamma_g+\gamma_s)/4$ is shown by the blue triangles in Fig. 4(e)). Further, even the more demanding criteria, $g \geq (\gamma_g+\gamma_s)/2$, is satisfied for $r_{slot} \lesssim 1.5$ nm ($(\gamma_g+\gamma_s)/2$ is shown by red circle dots in Fig. 4(e)). These results confirm that the extremely confined picopatch mode and the nanocavity mode are strongly coupled. [65,73]

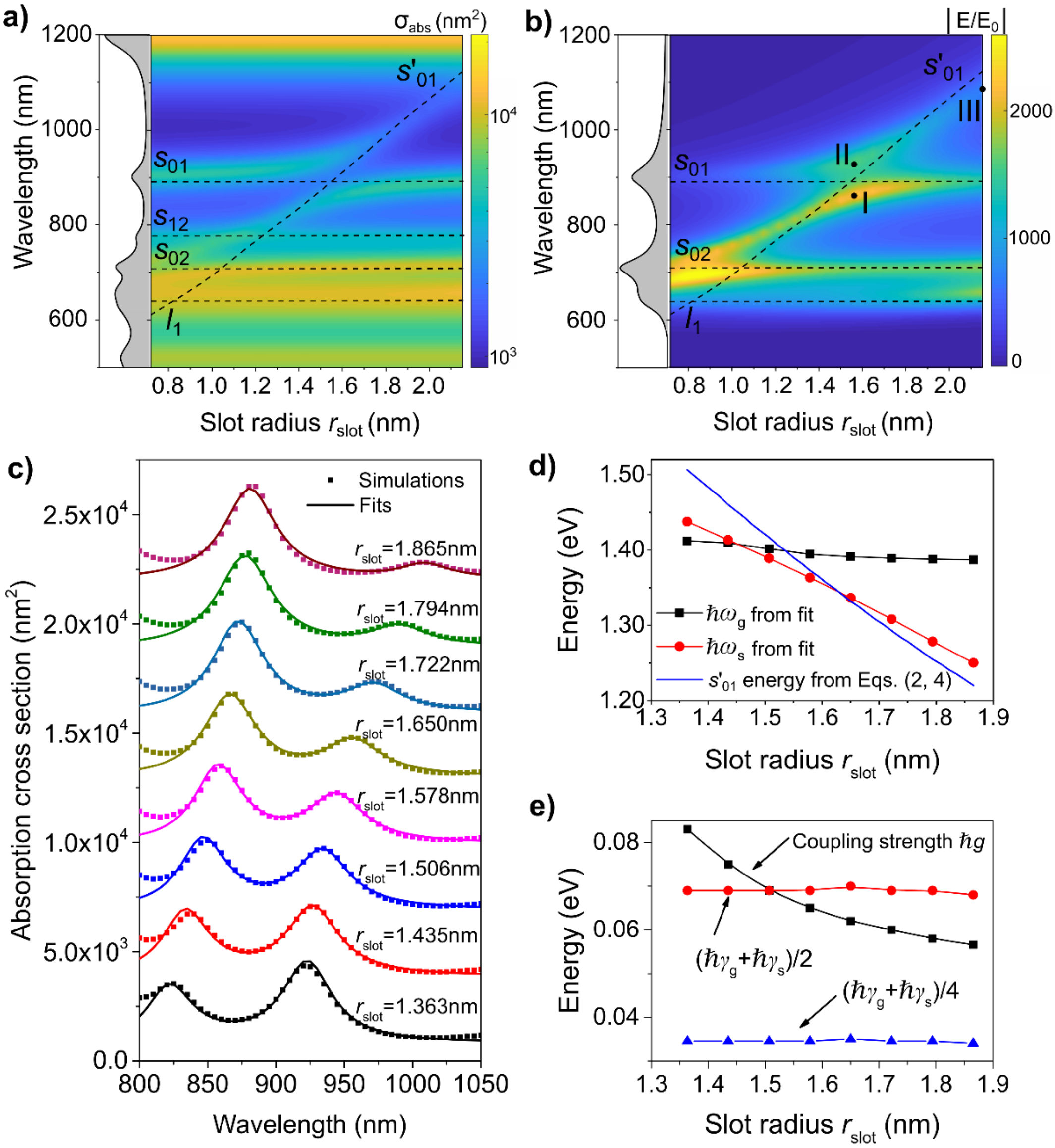

**Fig. 4 Mode hybridization and fitting based on the two coupled-oscillator model.** (a, b) Color maps of (a) the absorption cross section (in $nm^2$), and (b) the electric field enhancement $|E/E_0|$ (evaluated in the center of the slot formed between the metal substrate and the thin metal layer of the picopatch, marked by a black cross in Fig. 1(c)). These color maps are calculated as a function of the wavelength $\lambda$ and the slot radius $r_{slot}$ (sketch in Fig. 1(c)). The spectra to the left of each color map are the corresponding results for the bare NPoM nanocavity, consisting of the same system after removing the picopatch. The horizontal dashed lines indicate the position of the bare NPoM nanocavity eigenmodes. The theoretical wavelength of the $s'_{01}$ mode introduced by the picopatch is shown by the diagonal dashed line, as obtained from Eqs. (2) and (4) that do not include coupling between modes. Labels 'I', 'II', and 'III' correspond to conditions to be examined more in detail in Fig. 5. (c) Fittings of the absorption spectra obtained for $r_{slot}$ = 1.363 -1.865 nm and wavelength between $\lambda$ = 800 - 1050 nm. The fitting uses Eq. (5), derived from a coupled harmonic oscillator model, as discussed in the text. The dots correspond to the simulations, and the solid lines to the fits. Note that absorption spectra are shifted vertically for clearer visualization. (d) Resonant energies of the $s_{01}$ ($\hbar\omega_g$, black square dots) and $s'_{01}$ ($\hbar\omega_s$, red circle dots) modes

extracted from the fits for different values of $r_{slot}$, together with the resonant energy of the $s'_{01}$ mode predicted by Eq. (2) and Eq. (4) (blue solid line; these results are the same as those given by the diagonal dashed line in (a) and (b)). (e) Coupling strength $\hbar g$ between the $s'_{01}$ and $s_{01}$ modes (black square dots) and damping rates of the $s'_{01}$ and $s_{01}$ modes (red circle dots: $\hbar(\gamma_g+\gamma_s)/2$; blue triangles $\hbar(\gamma_g+\gamma_s)/4$) extracted from the fits for different values of $r_{slot}$.

### 4.4. Near-field distribution and effective mode volume

To complete our analysis of the picopatch and NPoM nanocavity modes and their coupling, we show in Fig. 5 the spatial distribution of the normalized amplitude of the electric field ($|E/E_0|$), i. e. field enhancement maps, of several relevant modes (plotted in the vertical x-z and horizontal x-y planes). Figures 5(a, b) correspond to the uncoupled mode of the bare NPoM nanocavity at $\lambda = 895$ nm (see red dashed lines in Figs. 3(a-c)). The fields and associated surface charges induced at the interfaces (the latter denoted schematically by '+' and '-') show an almost rotationally-symmetric standing-wave pattern with local maximum near the nanogap edges and strongest fields in the central region of the nanogap, which confirms the assignment of this peak as the $s_{01}$ TCP mode of the nanocavity.

To analyze the uncoupled $s'_{01}$ picopatch mode, we plot in Figs. 5(c, d) the field enhancement $|E/E_0|$ and corresponding charge distributions near the picopatch, for slot radius $r_{slot}$ = 2.153 nm and wavelength $\lambda$ = 1080 nm (conditions marked with 'III' in Fig. 4(b)). These conditions are chosen to minimize the interaction of the $s'_{01}$ with the $s_{01}$ nanocavity mode, due to a large detuning between the frequency of the $s_{01}$ and $s'_{01}$ modes. The fields, which are again rotationally symmetric, show extremely large E-field enhancement ($|E/E_0| \approx 700$) confined to the tiny vacuum slot in the middle of the picopatch. The field distribution of the $s'_{01}$ picopatch mode (Figs. 5(c, d)), with maxima at the slot center and weaker fields at the slot edges, differs from that of the $s_{01}$ nanocavity mode (Figs. 5(a, b)) due to the different boundary conditions at the edges of the guides (closed and open, respectively, see discussion of Eqs. (3-4)). Higher-order picopatch modes ($s'_{11}$ and $s'_{02}$) are discussed in section S5 of Supplementary Information (see Fig. S7).

The fields of the new hybrid modes emerging from the interaction of the $s_{01}$ NPoM nanocavity and $s'_{01}$ picopatch modes are calculated in the region of the anti-crossing ($r_{slot}$ = 1.578 nm, and λ = 865 nm or λ = 935 nm, marked with 'I' and 'II' in Fig. 4(b)) and plotted in Fig. 5(e) ($\lambda$ = 865 nm) and Fig. 5(f) ($\lambda$ = 935 nm). The associated charges induced at the metal-dielectric interfaces are again indicated schematically. The hybrid-mode fields combine features from the two original uncoupled modes, with the strongest fields located in the picopatch slot (see zooms in the right panels in (e, f)) but also strong fields at the center and near the edges of the nanocavity. Consistent with hybridization theory,[74,75] these electric field distributions can be understood as the sum of the contributions from the $s_{01}$ and $s'_{01}$ modes, with a phase difference between these contributions that depends on the specific hybrid mode: relative phase 0 for hybrid mode 'I' and π (sign change) for hybrid mode 'II', corresponding to symmetric and anti-symmetric combinations of the uncoupled plasmon modes. The relative phase can be appreciated in the figures by comparing the sign of the charges at the edges of the nanogap (set by the $s_{01}$ mode contribution) with those at the picopatch (dominated by the $s'_{01}$ mode contribution). The sign at the gap edges is the same in Figs. 5(a, e, f), but the sign at the picopatch is opposite in the hybrid mode 'II' (Fig. 5(f)) compared to the hybrid mode 'I' (Fig. 5(e)) and the uncoupled picopatch mode (Fig. 5(c)). For completeness, note that the charges at the center of the top surface of the nanocavity – 0.665 nm over the picopatch – are set mainly by the uncoupled $s_{01}$ nanocavity mode, which explains why they have the same sign for both hybrid modes.

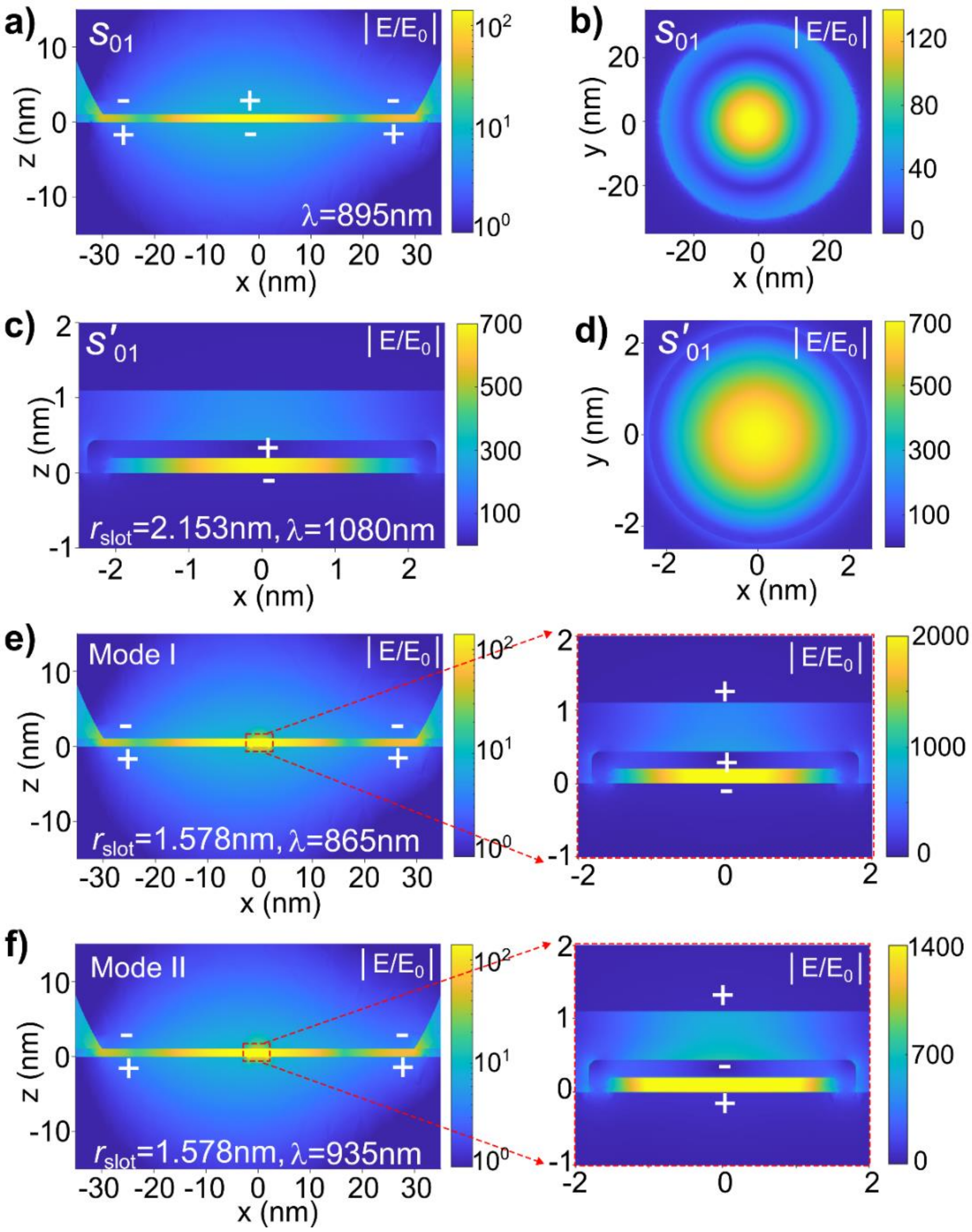

Fig. 5 **Near-field spatial distributions of different nanocavity and picopatch modes**. (a, b) Field enhancement ($|E/E_0|$) maps of uncoupled nanocavity mode $s_{01}$ (obtained at $\lambda$ = 895 nm for the bare NPoM without picopatch). (c, d) Field enhancement ($|E/E_0|$) maps of the (almost) uncoupled picopatch mode $s'_{01}$ (obtained at $\lambda$ = 1080 nm for the NPoM containing a picopatch with $r_{slot}$ = 2.153 nm; condition marked by dot with label “III” in Fig. 4(b)). (e, f) Field enhancement ($|E/E_0|$) maps of the hybrid modes marked “I” and “II” in Fig. 4(b) (obtained at $\lambda$ = 865 nm and $\lambda$ = 935 nm, respectively, for the NPoM containing a picopatch with $r_{slot}$ = 1.578 nm). (a, c, e, f) show the fields in the vertical x-z plane that contains the center of the nanogap (y=0) (side view). (b, d) show the fields in the horizontal x-y plane (top view), corresponding to (b) z = 0.55 nm (crossing the nanogap center) and (d) z = 0.1 nm (crossing the slot center). The left panels in (e, f) show the fields in the nanogap, and those on the right a zoom into the region of the picopatch. Notice that the region plotted in (c, d) is significantly smaller than in (a, b) (see the x and y labels). The colormap is logarithmic in (a) and left panels of (e, f) and linear in other figures (including right panels of (e, f)). The x and z directions are indicated in the axis and in Fig. 1(c), with the center of the nanogap corresponding to x=y=0, z=0.55 nm.

Additionally, the field distribution highlights that the fields in the vacuum slot of the picopatch are substantially enhanced when the hybrid modes are excited, with the enhancement factors reaching up to $|E/E_0| \approx 2000$. The fields can also significantly penetrate inside the metal, reaching $|E/E_0| \approx 300$ at the picopatch thin metallic layer (Au monolayer), which is consistent with recent work suggesting that the field increase associated with the random formation of picopatches could explain short-lived experimental increases (flares) of the inelastic light emission from gold nanogaps.[34-35] From the opposite perspective, it may be possible to use the flares to assess the experimental field enhancement in the picopatch. Last, the electric fields at the top of the picopatch, i.e., the region between the nanocavity upper interface and the gold monolayer below, are significantly smaller than in the slot but still very large (up to $|E/E_0| \approx 600$, compared to $|E/E_0| \approx 200$ in the case of the bare NPoM resonator; see also Fig. 3(c)). Since this region is available for molecules or emitters, the large field enhancement could therefore be useful in molecular spectroscopy, and the enhancement of the spectroscopic signal may also introduce a path to calibrate the field enhancement.[76,77]

The extremely large field enhancement and field concentration at the picopatch slot suggest that the effective mode volume ($V_{\mathrm{eff}}$) of the hybrid modes is very small. Here, $V_{\mathrm{eff}}$ refers to the volume that determines the local density of states (LDOS $\propto 1/V_{\mathrm{eff}}$) of a given mode and thus how much the mode can enhance different light-matter interactions in a quantum mechanical description, and not just to the volume of a 'hot spot' region of strong fields. To reach very small $V_{\mathrm{eff}}$ is significantly more challenging than obtaining 'hot spots' of the same volume. $V_{\mathrm{eff}}$ was initially introduced by Purcell[78] and has been typically calculated using the electromagnetic energy.[79,80] However, this approach can be inappropriate when losses are large, as in plasmonic systems,[81-84] and an alternative approach based on quasi-normal modes has been developed to treat non-Hermitian (lossy) systems.[85-88] In this case, the effective volume $\tilde{V}$ is a complex-valued volume and can be obtained from the following equation,

$$\tilde{V} = \frac{\int \left[ \left( \frac{\partial(\omega\tilde{\varepsilon})}{\partial\omega} \right) \tilde{\boldsymbol{E}}_s{}^2 - \mu_0 \tilde{\boldsymbol{H}}_s{}^2 \right] d^3\boldsymbol{r}}{2\varepsilon_0 \left( \tilde{\boldsymbol{E}}_m \cdot \boldsymbol{u} \right)^2}, \tag{6}$$

with $\tilde{\varepsilon}$ the position-dependent complex permittivity, $\varepsilon_0$ the vacuum permittivity, $\mu_0$ the vacuum permeability, $\omega$ the (angular) frequency and $\tilde{\boldsymbol{E}}_s$ ($\tilde{\boldsymbol{H}}_s$) the position-dependent complex scattered electric (magnetic) field vector of the resonant mode under analysis. Note that Eq. (6) takes into account that the slot is filled by vacuum, and that the integral extends over the whole computational domain. The complex Au permittivity (and its derivative) and the fields are evaluated at the corresponding resonant frequency. $\tilde{\boldsymbol{E}}_m \cdot \boldsymbol{u}$ is the value of $\tilde{\boldsymbol{E}}_s$ at a given position projected to a unit vector $\boldsymbol{u}$, so that in the quasi-normal modes formalism $\tilde{V}$ is defined for a given location and orientation. Here, we choose the position to be the center of the slot inside the picopatch and $\boldsymbol{u}$ = (0, 0, 1) (z-axis), which corresponds (up to an excellent approximation) to the position of strongest field and to the orientation of this field. We make this choice because the resulting $V_{\mathrm{eff}} = \mathrm{Re}\{\tilde{V}\}$ (Re{} indicating the real part) describes the field confinement in an analogous manner to the often-used convention in the analysis of normal modes where the mode volume is an intrinsic property that quantifies the confinement of a mode, without needing to specify any particular position.[81,89-91] The imaginary part of the effective volume, $\mathrm{Im}\{\tilde{V}\}$, quantifies the losses of the system.

All the quantities involved in Eq. (6), e.g. $\tilde{\boldsymbol{E}}_s$ and $\tilde{\boldsymbol{H}}_s$, are in principle evaluated at the complex-valued resonant frequency of the quasi-normal modes.[86] However, for simplicity, we fix the (real) frequency from the maxima in the absorption spectra, and obtain the scattered fields at this frequency under plane wave illumination. As the effective mode volume in Eq. (6) is defined for a single mode, in these calculations we diminish *ad-hoc* the imaginary part of the Au permittivity by a factor of 10 (i.e., $Im\{\varepsilon_{Au}\} \rightarrow Im\{\varepsilon_{Au}\}/10$), which should not modify noticeably the value of $\mathrm{Re}(\tilde{V})$ but avoids the difficulties due to the off-resonant excitation of other modes.

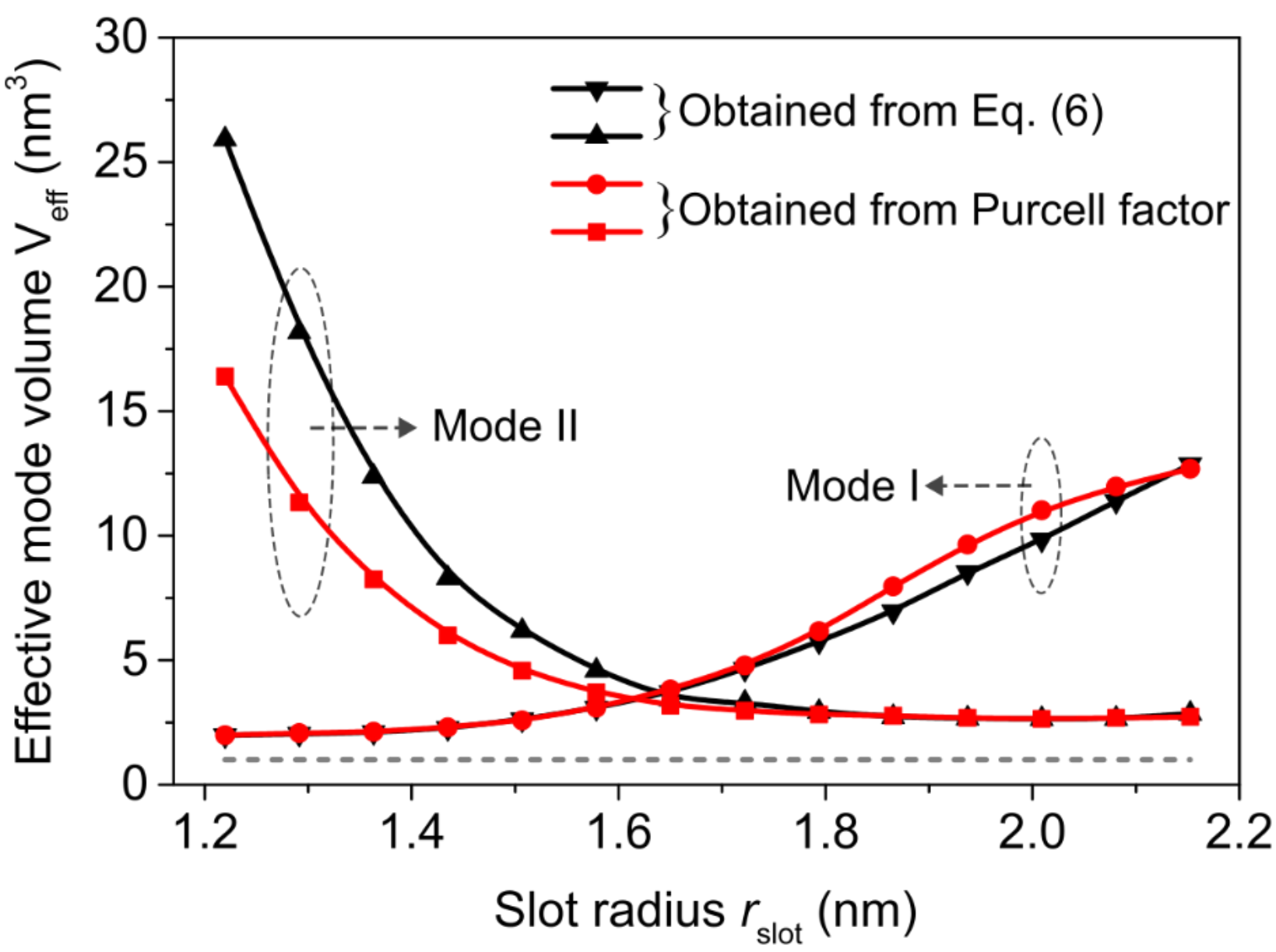

Fig. 6 **Effective mode volume $V_{eff}$ of the two hybrid modes of the NPoM with picopatch.** $V_{eff}$ as a function of the picopatch radius $r_{slot}$, obtained from the real part of Eq. (6) (black lines) and from the Purcell factor (red lines). The reference value of mode volume of 1 nm$^3$ is indicated by the horizontal gray dashed line. The simulated system corresponds to the NPoM resonator with picopatch sketched in Fig. 1(c).

Figure 6 shows the value of $V_{eff}$ = Re$\{\tilde{V}\}$, which describes the field confinement, as a function of $r_{slot}$ for the two resonant hybrid modes (I and II) of the NPoM resonator with picopatch (black lines). At $r_{slot}$ ≈ 1.650 nm, the field confinement of both hybrid modes is similar and extremely small, ~ 3.5 nm$^3$. This value of $r_{slot}$ is close to the value at which the frequencies of the uncoupled $s_{01}$ and $s'_{01}$ modes of the bare NPoM nanocavity and of the picopatch are the same in Fig. 4(d). As $r_{slot}$ increases or decreases from 1.650 nm, the mode volume of one of the hybrid modes strongly increases, while the other decreases gradually. This tendency follows the general expectation for the hybridization of the modes of very different effective mode volumes: as the detuning increases, the values of $V_{eff}$ approach towards those of the uncoupled modes. In our case, the hybrid mode with smaller mode volume is mostly associated with the picopatch $s'_{01}$ mode. For $r_{slot}$ ≈ 1.2 nm, the (real part of the) effective mode volume of this mode is only $V_{eff}$ ≈ 1.8 nm$^3$, similar to the geometrical volume of the slot ($V_{slot}$ ≈ 0.9 nm$^3$). We thus find that the effective mode volume of the NPoM resonator with picopatch can get close to 1 nm$^3$ (value marked for reference by the horizontal dashed line in Fig. 6), and may become even smaller than this value for more optimized picopatch configurations, thus offering an excellent geometrical alternative to achieve extreme light confinement to the picocavities formed by isolated atomic-sized protrusions.[23]

In order to further verify the value of $V_{\mathrm{eff}}$ obtained in Fig. 6 with Eq. (6) and real-valued frequencies, we also extract the effective mode volume from the Purcell factor[78] experienced by a dipole positioned at the center of the picopatch slot and perfectly aligned along the vertical z-axis. In section S6 of the Supplementary Information, we describe this procedure in detail and plot the Purcell factor as a function of wavelength and slot radius (Fig. S8). The effective mode volume for the two hybrid modes (I and II) of the NPoM resonator with picopatch, obtained from the Purcell factor, is plotted in Fig. 6 (red lines). The values of the effective mode volume obtained in this way are in good overall agreement with the values of $V_{\mathrm{eff}}$ obtained from Eq. (6). For reference, we also show in Fig. S11 in the Supplementary Information the Purcell factor at a different position in the gap. Last, as an additional verification of the validity of using real-valued frequencies to calculate $V_{\mathrm{eff}} = \mathrm{Re}\{\tilde{V}\}$ with Eq. (6), we have also calculated this quantity for $r_{\mathrm{slot}}$ = 1.435 nm by evaluating Eq. (6) with complex frequencies according to the full quasi-normal mode approach.[85,86,88] The values of $V_{\mathrm{eff}}$ obtained with the real and complex frequencies are almost identical, as discussed in more detail in section S10 of the Supplementary Information.

### 4.5. Robustness to morphology and losses

The picopatch that we have considered forms a vacuum slot of perfectly homogeneous thickness. However, the exact morphology of a picopatch cannot be controlled in current experiments, where the picopatch is formed through the random movement of gold atoms. To illustrate the influence of the shape of the picopatch on the plasmonic response of the system, we consider next the picopatch configuration sketched in Fig. 7(a). In this case, the top flat surface of the picopatch is replaced by a gold semi-ellipsoid of wall thickness $t$ = 0.235 nm, so that a non-flat slot with a maximum height $\delta$ = 0.2 nm and a radius at the bottom $r_{slot}$ is formed below. Other than this modification, the rest of geometrical properties of the NPoM resonator with picopatch remain unchanged.

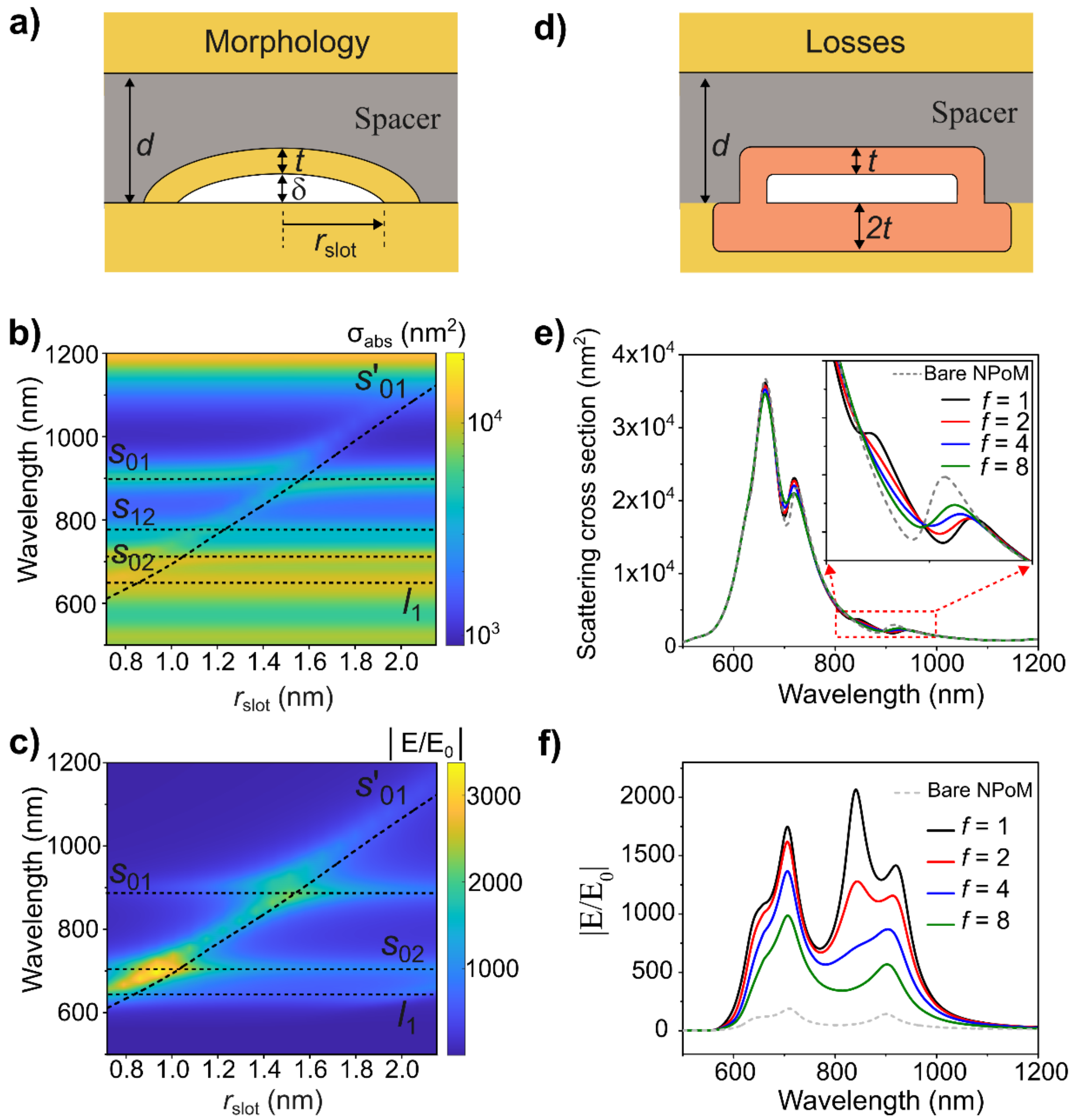

**Fig. 7 Robustness of the response to the exact shape of the picopatch and to material losses.** (a) Sketch (not to scale) of a modified picopatch morphology, where the top surface of the picopatch corresponds to a rotationally-symmetric semi-ellipsoid gold monolayer of thickness $t$ = 0.235 nm with short semi-axis (height) of size $\delta$ = 0.2 nm and long semi-axis (radius) of size $r_{slot}$. (b, c) Color maps of the (b) absorption cross section (in nm$^2$) and (c) E-field enhancement ($|E/E_0|$) spectra, for a NPoM resonator presenting the modified picopatch in (a) and otherwise the same geometry as in Fig. 1(c). The field enhancement is obtained at the center of the slot, and the colormaps show the dependence of the spectra on $r_{slot}$. The resonant wavelength of the NPoM nanocavity eigenmodes (which are independent of $r_{slot}$), are indicated by horizontal dashed lines in the color maps. The theoretical evolution of the uncoupled $s'_{01}$ picopatch mode with $r_{slot}$ (given by Eqs. (2) and (4)) is indicated by the diagonal dashed line in the color maps. (d) Sketch (not to scale) of the picopatch region of a resonator, showing (rotationally symmetric) regions marked with reddish

color where the permittivity of Au is changed through the loss factor *f,* following Eq. (7). (e) Scattering and (f) electric field enhancement $|E/E_0|$ spectra as a function of the loss factor *f* in the picopatch (with increasing *f* corresponding to larger absorption losses), for a resonator of otherwise the same geometry as in Fig. 1(c) ($r_{slot}$ = 1.435 nm). The field enhancement is obtained at the center of the vacuum slot in the picopatch (marked by a black cross in Fig. 1(c)). The inset in (e) shows a zoom-in view of the scattering spectra from 800 nm to 1000 nm. Spectra of the same system but without the picopatch (bare NPoM nanocavity) are plotted as reference by the gray dashed line. In this case, the much weaker field enhancement is evaluated in the center of the nanogap.

The absorption cross section spectra of the modified system with a non-flat picopatch and the corresponding electric field enhancement $|E/E_0|$ spectra (evaluated in the slot center) are shown by color maps in Figs. 7(b, c) as a function of $r_{slot}$. A clear anti-crossing-like behavior is observed when the $s_{01}$ mode associated with the bare NPoM nanocavity (marked by a horizontal dashed line) is resonant with the $s'_{01}$ picopatch mode (the diagonal dashed line indicates the resonant wavelength of this mode in the absence of coupling with other modes, as given by Eq. (2) and Eq. (4)). Crucially, the qualitative behavior remains the same as for the flat picopatch in Figs. 4(a, b), although a more careful comparison between Figs. 4(a, b) and Figs. 7(b, c) shows some differences. For example, the anti-crossing behavior is less clear for the semi-elliptical picopatch, with a smaller energy difference between the two hybrid modes, indicating a smaller coupling strength than the case in Figs. 4(a, b). Further, the maximum E-field enhancement reaches ~ 3000 for the semi-elliptical picopatch, even larger than for the flat picopatch. We thus find that the exact plasmonic response of the full system is sensitive to the geometrical details of the picopatch morphology, but the general behavior remains unchanged. For completeness, the scattering spectra of the modified system with a semi-elliptical picopatch is shown in Fig. S9 of Supplementary Information. We also note that changes of the picopatch morphology that break the rotational symmetry of the system would lift the degeneracy of some of the cavity modes and could lead to mode splitting also in the absence of strong coupling.[92]

Last, we consider the effect of increased material losses on the response of the system. We discussed previously how quantum calculations confirm the existence of the slot mode in the MIMI configuration even for extremely thin layers, but the influence of quantum effects on the losses experienced by the slot mode is more challenging to assess. Nevertheless, in principle, we expect that the absorption in gold can be larger than that predicted from the classical permittivity due to non-local, and spill-out effects. To analyze to what extent the optical response of the system is sensitive to an increase in absorption losses, we artificially increase the imaginary part of the Au classical permittivity according to the expression,

$$\varepsilon_{Au} = Re\{\varepsilon_{Au}\} + ifIm\{\varepsilon_{Au}\}, \tag{7}$$

where the loss factor is $f$ = 1 for the experimental values in Ref. [39] (i.e. the permittivity used in other simulations in this work when not stated otherwise) and $f$ > 1 corresponds to increased absorption losses. As the increase in absorption losses is likely local, we adopt this modified permittivity only in the regions around the vacuum slot, i.e., the lifted Au monolayer that forms the slot and the 0.47 nm-thick top of the Au substrate just below (corresponding to ≈2 atomic layers), as sketched by the reddish regions in Fig. 7(d).

The simulated scattering spectra for increasing loss factor $f$ are plotted in Fig. 7(e) (solid lines). In general, the scattering spectrum depends weakly on the loss factor $f$, which could be expected as the effect of the presence of the picopatch was already relatively small (Fig. 3(a)). However, the NPoM with picopatch spectra remains different from that of the reference bare NPoM nanocavity without picopatch (gray dashed line) in the spectral region around $\lambda \approx$ 800 - 1000 nm (see zoom in inset of Fig. 7(e)), even for strongly increased losses (e.g., $f$ = 8, green line).

The effect of the increased losses is clearer in the field enhancement ($|E/E_0|$) spectra, as shown in Fig. 7(f). As the loss factor $f$ is increased, $|E/E_0|$ (calculated in the center of the picopatch slot) decreases substantially over the whole spectral range. Furthermore, in the spectral region around $\lambda \approx$ 800 -1000 nm, the two peaks associated with the hybrid modes become just one as $f$ is increased. However, the field enhancement factor remains very large for all values of $f$ considered, several times larger than those obtained for the bare NPoM nanocavity without the picopatch (gray dashed line). Thus, we find that even when the local losses are increased to be 8-fold larger than the standard value, the strong field enhancement that makes picopatches interesting for nanophotonics remains robust. Further, we show in section S8 of the Supplementary Information that this finding applies not only to the field enhancement at the center of the picopatch slot, but also near the center of the nanocavity.

## 5. Conclusion

In summary, we have analyzed in detail the optical response of NPoM resonators containing picopatches that are formed when an atomic-thick section of atoms in a NPoM nanocavity is lifted from the top layer of the underlying metallic substrate, leaving a sub-nanometer-thick vacuum slot below. This situation was identified in recent work as a morphology that could explain flare effects in metal nanoparticle fluorescence.[34] Using classical electromagnetic calculations (justified by previous quantum atomistic calculations of infinite MIMI waveguides[33]), we show that the slot mode can be tuned over a broad wavelength range and that it hybridizes with the bare NPoM cavity modes when both modes overlap spectrally. The hybridized mode has an effective mode volume close to 1 $nm^3$, and induces a huge ≈ 2000-fold electric field enhancement. Furthermore, these fields can penetrate into the metal forming the picopatch, and thus potentially explain emission flares observed in experiments. The picopatches also affect the far-field spectra, which might open an alternative path to detect their formation. Last, we illustrate how these general results are robust to changes in the picopatch shape or quantum-induced (non-local) increases of absorption losses. Thus, this study shows that the presence of picopatches in nanoresonators is a promising configuration to achieve extreme field localization and enhancement in nanophotonics.

**Acknowledgements**

We thank Philippe Lalanne for his help with the quasi-normal mode formalism. We

acknowledge grant PID2022-139579NB-I00 funded by MICIU/AEI/10.13039/501100011033 and by ERDF, EU, and grant no. IT 1526-22 from the Department of Science, Universities and Innovation of the Basque Government. J. H. and Y. Z. acknowledge funding from grants 17704208, 62475242 funded by National Natural Science Foundations of China, grant 2024YFE0105200 from National Key R&D Program of China, grant 252300420342 funded by Natural Science Foundation of Henan Province. X. A. acknowledges Spanish Ministerio de Ciencia, Innovación y Universidades for his PhD grant no. FPU21/02963.

**Note:** This article is under review in Laser & Photonics Reviews.

# Supplementary Information

## Contents

## S1. Dispersion of MIM and MIMI plasmons

We derive in this section the dispersion relationship of the laterally-infinite structures that are used in the main text to analyze the response of the NPoM resonators with and without picopatch. These laterally-infinite structures form plasmonic waveguides that can guide surface plasmon polaritons.

### S1.1 Dispersion of the MIM plasmons

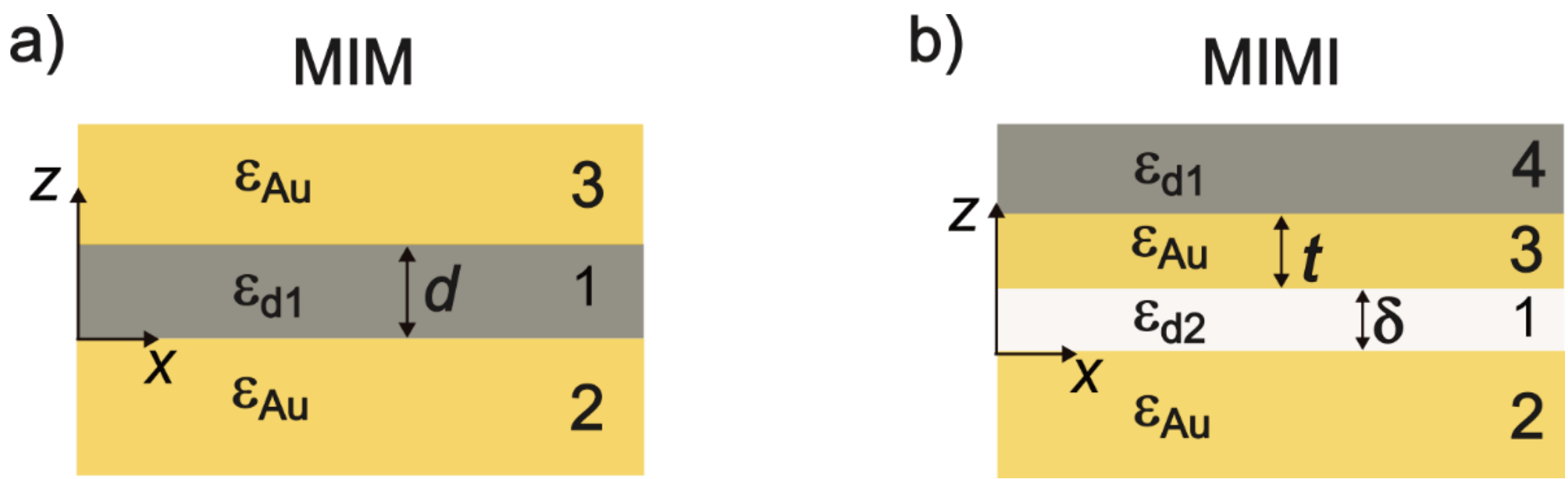

**Fig. S1 Sketch of MIM and MIMI planar waveguides.** (a) MIM structure consisting of a dielectric layer (permittivity $\varepsilon_{d1}$) of thickness *d* that separates two semi-infinite gold layers (permittivity $\varepsilon_{Au}$). The whole space is thus divided into three regions, denoted with "1" (dielectric of permittivity $\varepsilon_{d1}$), and "2" and "3" (both semi-infinite Au layers). (b) MIMI structure consisting of a thin gold layer of thickness t separated by a vacuum layer of thickness δ (permittivity $\varepsilon_{d2}$) from a gold semi-infinite layer. On top of the thin gold layer, there is a semi-infinite dielectric layer (permittivity $\varepsilon_{d1}$). The four regions are denoted with "1" (dielectric of permittivity $\varepsilon_{d2}$), "2" (bottom Au semi-infinite substrate), "3" (thin upper gold layer) and "4" (dielectric of permittivity $\varepsilon_{d1}$). The z-direction is perpendicular to the interfaces shown in (a-b), with z=0 at the top surface of the bottom gold layer, as indicated by the plotted axis.

We first consider the infinite planar Metal-Insulator-Metal (MIM) multilayer waveguide sketched in Fig. S1(a), composed of two Au semi-infinite substrates of permittivity $\varepsilon_{Au}$ separated by a dielectric (or equivalently, insulator) layer of thickness *d* and permittivity $\varepsilon_{d1}$. The dispersion of the plasmonic waveguide modes can be obtained, for example, by using the wave-transfer matrix formalism or considering the infinite reflections of a plane wave in the two dielectric-metal interfaces,[1-3] and searching for the pole of the resulting expression. According to this procedure, the dispersion can be obtained directly by solving the equation

$$1 - r_{12}r_{13}e^{2ik_{1z}d} = 0, \quad \text{(S1.1)}$$

where $r_{12}$ = $r_{13}$ are the reflection coefficients of plane waves propagating in the dielectric layer (region "1"), at the interface with the Au semi-infinite regions (regions "2" and "3"). $k_{iz}$ is the $z$ component of the wavevector of a plane wave in the medium "i". The reflection coefficients are given by the general Fresnel equations,

$$r_{ij} = \frac{\varepsilon_j k_{iz} - \varepsilon_i k_{jz}}{\varepsilon_j k_{iz} + \varepsilon_i k_{jz}}. \quad \text{(S1.2)}$$

In our case, $\varepsilon_1 = \varepsilon_{d1}$, $\varepsilon_2 = \varepsilon_3 = \varepsilon_{Au}$ in Eq. (S1.1), and

$$k_{1\text{z}} = \sqrt{\varepsilon_{d1}{k_0}^2 - q^2}, \quad k_{2\text{z}} = k_{3\text{z}} = \sqrt{\varepsilon_{Au}{k_0}^2 - q^2}, \quad \text{(S1.3)}$$

with $k_0=2\pi/\lambda$ the wavevector of a plane wave in vacuum and $q$ the in-plane wavevector of the plasmonic waveguide modes. $q$ is the same for all layers. By inserting Eq. (S1.2) into Eq. (S1.1), and using the definition of $k_{1z}$ and $k_{2z}$ =$k_{3z}$ given by Eq. (S1.3), we obtain the well-known dispersion of the MIM waveguide modes,[4] corresponding to Eq. (1) in the main text,

$$\tanh\left(\frac{id}{2}\sqrt{{k_0}^2\varepsilon_{d1} - q^2}\right) = \frac{\varepsilon_{d1}}{\varepsilon_{Au}}\sqrt{\frac{q^2 - \varepsilon_{Au}{k_0}^2}{q^2 - \varepsilon_{d1}{k_0}^2}}\ . \quad \text{(S1.4)}$$

**S1.2 Dispersion of the MIMI plasmons**

We analyze next the Metal-Insulator-Metal-Insulator (MIMI) planar structure as sketched in Fig. S1(b). From bottom to top, this structure consists of a semi-infinite gold substrate (region "2" of permittivity $\varepsilon_{Au}$), a dielectric layer of permittivity $\varepsilon_{d2}$ of thickness $\delta$ (region "1"), a thin gold layer of thickness $t$ (region "3"), and a semi-infinite dielectric region of permittivity $\varepsilon_{d1}$ (region "4"). We follow the same approach as in Ref. [5] except that we do not assume $\varepsilon_{d1} = \varepsilon_{d2}$. The dispersion of the MIMI waveguide modes in the region "1" can be obtained from an expression similar to Eq. (S1.1),

$$1 - r_{12} r_{th} e^{2ik_{1z}\delta} = 0, \quad \text{(S1.5)}$$

where $r_{12}$ and $r_{th}$ represent the reflected coefficients of a plane wave propagating in region "1", at the interface with regions "2" and "3", respectively. $r_{th}$ takes into account the thin gold layer and the top dielectric material, not just the dielectric-gold interface. $k_{iz}$ indicates again the z component of the wavevector of the plane wave propagating in the region "i". The coefficient $r_{12}$ can be obtained from Eq. (S1.2) with $\varepsilon_1 = \varepsilon_{d2}$ and $\varepsilon_2 = \varepsilon_{Au}$. On the other hand, at the interface between the region "1" and the thin gold layer, the reflection coefficient is given by

$$\mathrm{r}_{th} = \frac{(r_{13} + r_{34} e^{2ik_{3z}t})}{1 - r_{34} r_{31} e^{2ik_{3z}t}}, \quad \text{(S1.6)}$$

as can be derived again by using the wave-transfer matrix formalism or by considering the infinite reflections of a plane wave illuminating a thin film embedded between two different media.[1-3] The reflection coefficient $r_{13}$ and $r_{34}$ are obtained from Eq. (S1.2) with $\varepsilon_1 = \varepsilon_{d2}$, $\varepsilon_4 = \varepsilon_{d1}$ and $\varepsilon_3 = \varepsilon_{Au}$.

Due to the presence of a gold and a dielectric layer of sub-nanometric thickness, the in-plane wavevector of the MIMI waveguide mode is very large. Thus, we use the approximation $k_{1z} \approx k_{2z} \approx k_{3z} \approx iq$ , with $q$ the parallel component of the wavevector of the MIMI mode. By inserting Eq. (S1.6) into Eq. (S1.5), replacing $k_{1z}$, $k_{2z}$, and $k_{3z}$ with $iq$, and looking for the solutions of the resulting expression (corresponding to the poles of the response of the full system), we obtain the equation whose solution gives us the dispersion of the MIMI plasmonic waveguide mode (Eq. (2) in the main text)

$$\left(\frac{\varepsilon_{Au} - \varepsilon_{d1}}{\varepsilon_{Au} + \varepsilon_{d1}}\right)\left(\frac{\varepsilon_{Au} - \varepsilon_{d2}}{\varepsilon_{Au} + \varepsilon_{d2}}\right)\left[\left(\frac{\varepsilon_{Au} + \varepsilon_{d1}}{\varepsilon_{Au} - \varepsilon_{d1}}\right)\left(\frac{\varepsilon_{Au} - \varepsilon_{d2}}{\varepsilon_{Au} + \varepsilon_{d2}}\right) - \left(\frac{(\varepsilon_{Au} + \varepsilon_{d1})(\varepsilon_{Au} - \varepsilon_{d2})}{(\varepsilon_{Au} - \varepsilon_{d1})(\varepsilon_{Au} + \varepsilon_{d2})} - e^{-2qt}\right)\left(1 - e^{-2q\delta}\right)\right] - 1 = 0 \quad (S1.7)$$

This equation is consistent with the one obtained in Ref. [5].

**S2. Further simulation details and convergence tests**

In this section, we present a description of the model implemented in COMSOL, and show convergence tests of the model. Details of the model are shown in Figs. S2(a, b). The NPoM resonator containing the picopatch is embedded in air (air regions are marked in gray, gold regions in yellow and dielectric in purple), forming the whole simulated domain with a radius $R_{dom}$. The simulated domain was surrounded by a sphere-shaped Perfect-Matched-Layer (PML) of thickness $T_{PML}$. Considering the model symmetry under the excitation condition (*p*-polarized plane wave at oblique incidence), we only include explicitly in the simulations half of the full model structure, as depicted in Figs. S2(a, b), by setting the boundary condition of "Perfect Magnetic Conductor (PMC)" for the cutting plane parallel to the x-z plane (normal to the direction of magnetic fields) that goes through the middle of the structure. This strategy reduces by half the number of elements of the mesh grids and consequently decreases the required computer resources (such as RAM memory) while maintaining the same mesh size. Notice that the simplified model is used just for the calculation of the scattering and absorption cross sections as well as the near-field of the proposed structure. When calculating the Purcell factor in section S6 by positioning a dipole at the center of the slot (without external excitation source), we adopt a full model structure due to the breaking of the symmetry characterizing the system under p-polarized illumination.

We mesh the implemented model with different mesh sizes for different regions, as shown in Figs. S2(c-e). In the air and gold substrate regions, the maximum and minimum element size of the mesh were set as Mp × 1/30 and Mp × 1/100 of the minimum incident wavelength $\lambda_{\min}$ considered (e.g., $\lambda_{\min}$ = 500 nm), respectively (here Mp is a factor controlling the meshing size up to convergence). For the region containing the gold nanosphere, the maximum and minimum size were set as Mp × $\lambda_{\min}$/150 and Mp × $\lambda_{\min}$/500, respectively. The limit above does not include the region inside and near the nanogap (in blue in Fig. S2(d)): in these regions we created a small cylinder of diameter equal to the width of the facet and height equal to the nanogap thickness, just below the gold sphere. The maximum and minimum element size for this region (excluding very close to the picopatch) were set as Mp × $\lambda_{\text{min}}$/1000 and Mp × $\lambda_{\text{min}}$/5000, and for the other regions in dielectric layer of permittivity $\varepsilon_{d1}$ (i.e., outside the nanogap) were Mp × $\lambda_{\text{min}}$/300 and Mp × $\lambda_{\text{min}}$/1000. Last, for the picopatch region (indicated by blue in Fig. S2(e)), we adopted an even finer element size, where the maximum and minimum mesh size were Mp × $\lambda_{\text{min}}$/3000 and Mp × $\lambda_{\text{min}}$/10000, respectively.

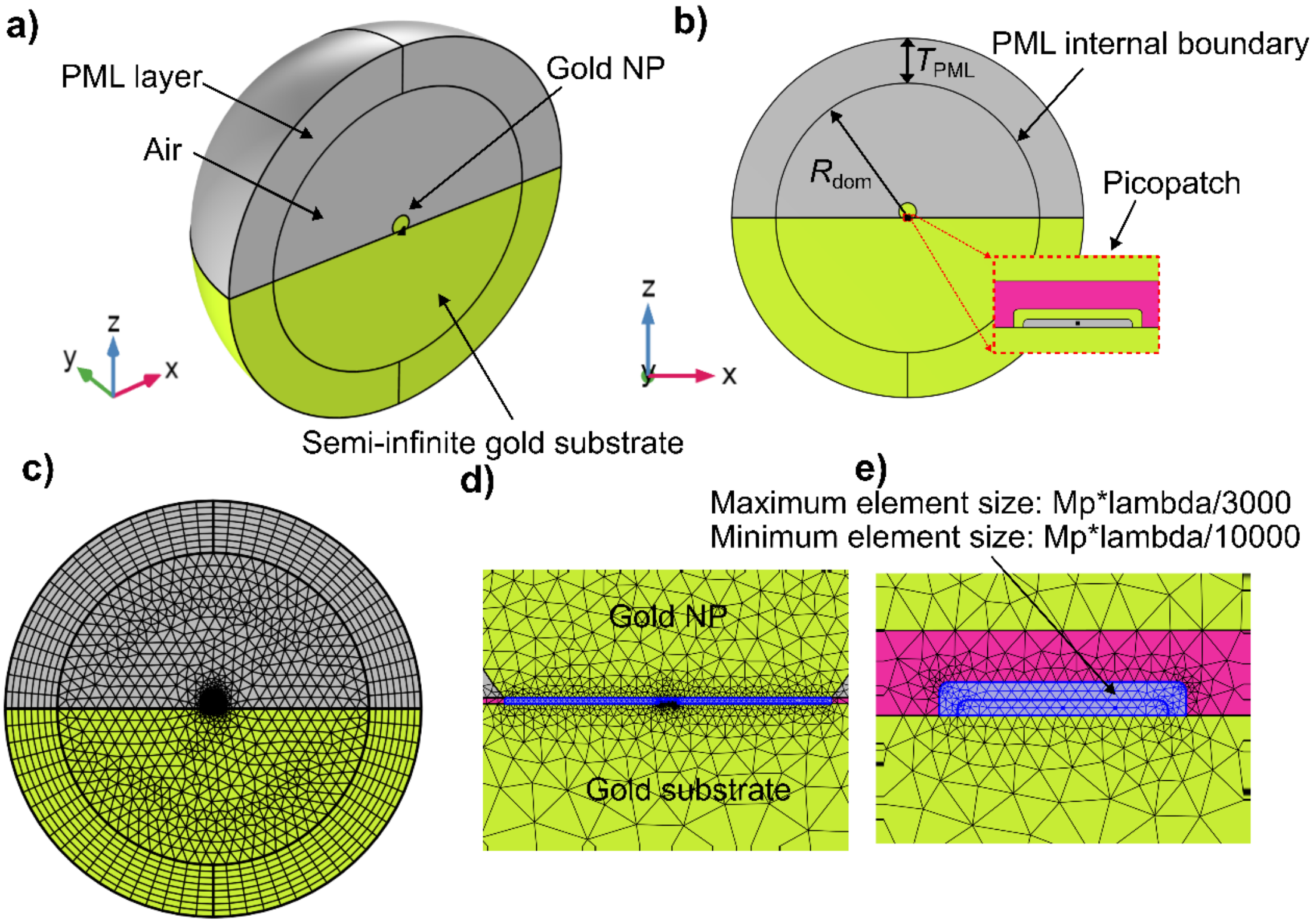

**Fig. S2 Setting-up of the model for the electromagnetic simulations.** (a) Geometry simulated with COMSOL, showing the NPoM with picopatch embedded in an air sphere region. (b) Vertical z-x cut of the geometry simulated. (c-e) Meshing of the geometry simulated (corresponding to that in Fig. 1(c) in the main text). The meshing size is different for different regions. For the picopatch region (indicated by blue in Fig. S2(e)), the maximum element size is Mp × $\lambda_{\text{min}}$/3000, and the minimum element size is Mp × $\lambda_{\text{min}}$/10000, where Mp is the factor controlling the meshing size, and $\lambda_{\text{min}}$ is the minimum incident wavelength (e.g., 500 nm). The blue color areas in (d) and (e) highlight the mesh in the gap and the picopatch respectively. In this figure, Mp=1.

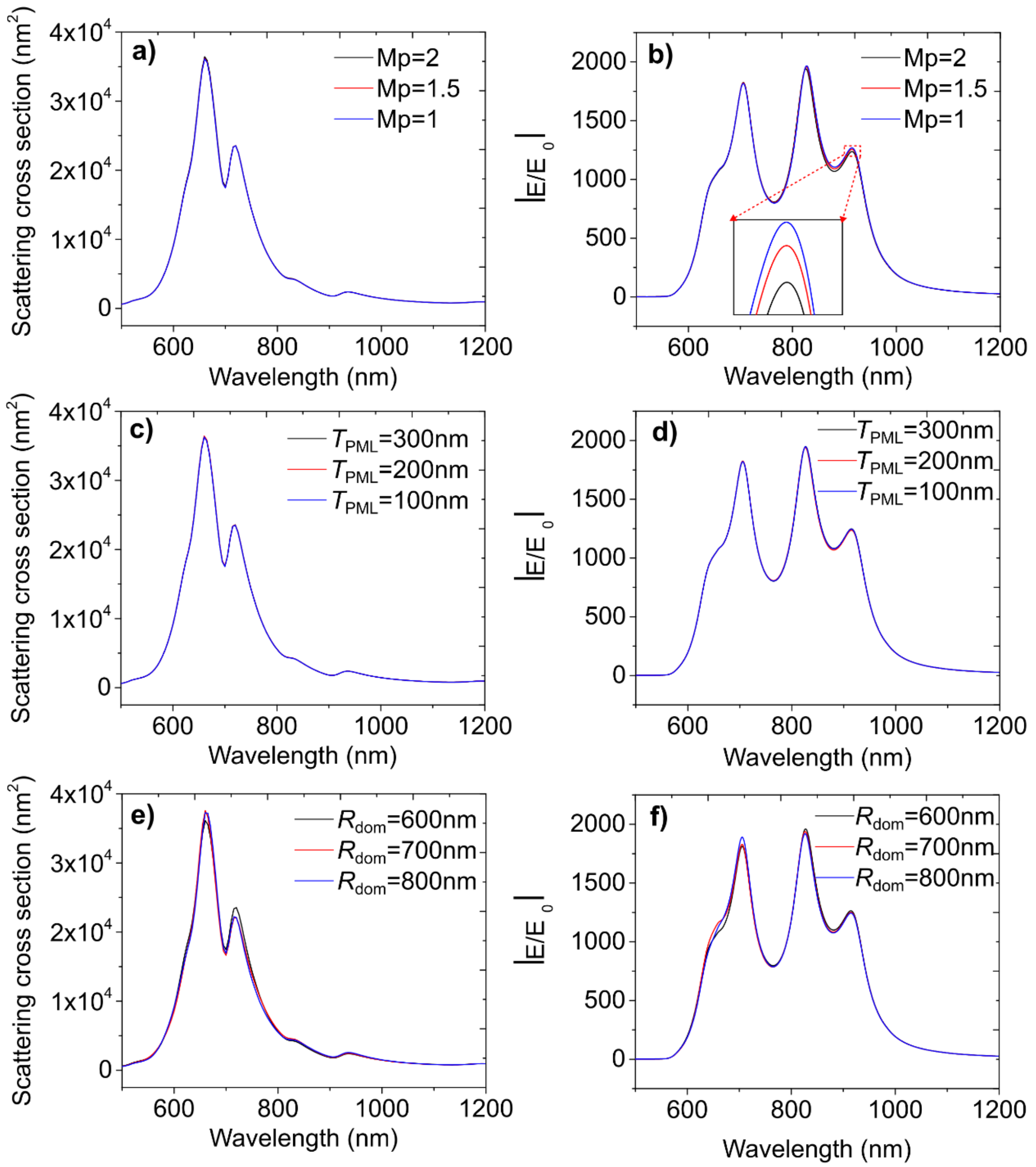

**Fig. S3 Convergence test of the electromagnetic simulations**. (a, c, e) Scattering cross section and (b, d, f) near-field enhancement $|\boldsymbol{E}/\boldsymbol{E_0}|$ spectra for (a, b) different meshing size, parameterized by Mp as indicated in the text, (c, d) different PML thickness and (e, f) different radius of the whole simulated domain. Results in (a-f) are obtained for the following geometrical parameters: $r_{\text{sphere}}$ = 40 nm, $r_{\text{facet}}$ = 30 nm, $d$ = 1.1 nm, $\delta$ = 0.2 nm, $t$ = 0.235 nm, and $r_{\text{slot}}$ = 1.435 nm. When not being changed, Mp = 2, $T_{\text{PML}}$ = 200 nm and $R_{\text{dom}}$ = 600 nm in this figure.

To test the convergence of the model, we simulated the scattering (Figs. S3(a, c, e)), and absorption (not shown) cross section spectra as well as the E-field enhancement $|\boldsymbol{E}/\boldsymbol{E_0}|$ (evaluated in the slot center, Figs. S3(b, d, f)) spectra, obtained for different values of the mesh grid size (Mp) (Figs. S3(a, b)), the PML layer thickness ($T_{PML}$) (Figs.

S3(c, d)) and the size of the simulated domain ($R_{dom}$) (Figs. S3(e, f)). The scattering cross section was obtained by computing the surface integral of the scattered Poynting vector ($\boldsymbol{P}_{sca}$) in the upper semi-sphere at the internal boundary of the PML layer, and then dividing by the incident laser intensity $I_0$ ($\sigma_{sca} = 2 \iint \boldsymbol{P}_{sca} \cdot d\boldsymbol{S} / I_0$). The absorption cross section was obtained from the volume integral of resistive losses ($Q_{\mathrm{rh}}$) in the metallic regions occupied by the gold nanosphere and the lifted gold monolayer, and then dividing by the incident laser intensity $I_0$ ($\sigma_{abs} = 2 \iiint Q_{rh} dV / I_0$). Notice that the factor of 2 in the two equations is introduced due to the use of the half of the full model structure, as described above. The results in Fig. S3 indicate that the convergence is excellent in all cases, with only small modifications of the spectra as we change Mp from 1 to 2 (Figs. S3(a, b)), the PML thickness $T_{PML}$ from 100 nm to 300 nm (Figs. S3(c, d)) or $R_{dom}$ from 600 nm to 800 nm (Figs. S3(e, f)). All the results in the main text and in the rest of Supplementary Information are obtained for the values of Mp = 2, $R_{\mathrm{dom}}$ = 600 nm and $T_{\mathrm{PML}}$ = 200 nm.

## S3. Assignment of NPoM nanocavity eigenmodes and near-field distributions

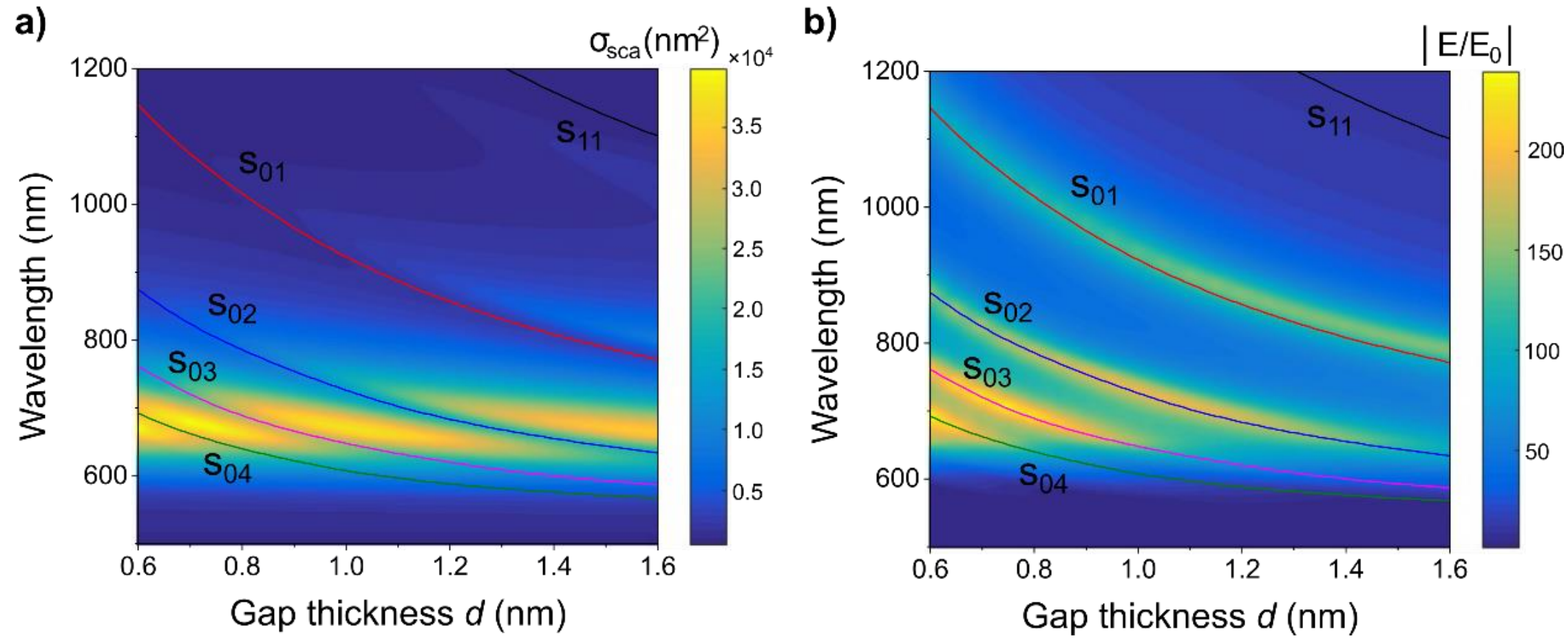

**Fig. S4 Analysis of the modes of the bare NPOM nanocavity.** (a) Scattering cross section and (b) E-field enhancement ($|E/E_0|$) spectra of the bare NPoM nanocavity (i.e., without picopatch) as a function of the gap thickness *d*. The geometry is otherwise the same as in the main text ($r_{sphere}$ = 40 nm, and $r_{facet}$ = 30 nm). The enhancement is obtained at the center of the gap of the NPoM nanocavity. The different lines superimposed to the color map correspond to the wavelengths of the NPoM TCP cavity modes according to Eq. (1) [equivalently, Eq. (S1.4)] and Eq. (3) in the main text.

We discuss in this section the modes of the bare NPoM nanocavity (i.e., the cavity without picopatch) in more detail. For this purpose, Fig. S4 shows the scattering and near-field enhancement $|E/E_0|$ spectra of the bare NPoM nanocavity as a function of the gap thickness *d*. The scattering is generally strong in the region around wavelength λ ≈ 650 - 700 nm, which corresponds to the excitation of the lowest energy Longitudinal Antenna Plasmon (LAP), $l_1$. This mode is associated with currents flowing along the vertical z direction (i.e., perpendicular to the Au surfaces in the gap, see axis and sketch in Fig. 1(c) in the main text). For flat facets, the LAP mode depends relatively weakly on the gap thickness for the values of *d* considered here.[6] Additionally, the scattering spectra also show a series of (comparatively) spectrally narrow minima and maxima that redshift strongly with decreasing *d*. We associate these features with the Transverse Cavity Plasmons (TCP) in the flat gap.

To estimate the resonant wavelength of these TCP modes, we use Eq. (3) and Eq. (1) in the main text (the latter corresponding to Eq. (S1.4)). These equations consider the gap as a Fabry-Pérot-like structure, with resonances emerging when the phase accumulated by a plasmon propagating parallelly to the flat facet at the gap and reflecting at the gap edges (i.e., at the limit of the flat facet in the x-y plane) is a multiple of 2π. The values obtained within this approach for modes $s_{mn}$ of different mode order (m, n), plotted as a function of the gap thickness *d,* correspond to the solid lines in Fig. S4. As discussed in more detail below and in the main text, the integers m and n are related to the spatial dependence of the field in the gap in the azimuthal and radial directions, respectively.

The predicted resonant wavelength of the $s_{0n}$ modes generally follow quite closely the minima in the calculated scattering spectra. A minimum is obtained here for these TCP modes instead of a maximum because of the destructive interference with the LAP mode ($l_1$) (and/or with a spectrally broad 'background' originating from higher-energy modes).[7,8] The $s_{11}$ mode does not exhibit any clear signature in the scattering spectra because it is weakly radiative, but can be identified in the absorption spectra (Fig. S5(a)). The good agreement between the analytical equations and the simulations is further substantiated in Fig. S4(b), which shows that the analytical prediction of the resonant wavelengths of the $s_{0n}$ modes follows the simulated maxima of the near-field enhancement. Here the $s_{11}$ mode is again not observable in the E-field enhancement spectra because this mode presents a zero at the gap center, the position where the E-field enhancement is evaluated. The overall excellent agreement with the calculations allows us to identify the (m, n) order of the different TCP modes, and strongly supports the validity of the analytical equations and the physical picture of the TCP modes as Fabry-Pérot-like resonances that originate from reflection of the MIM plasmons at the nanocavity edges.[7, 9-11]

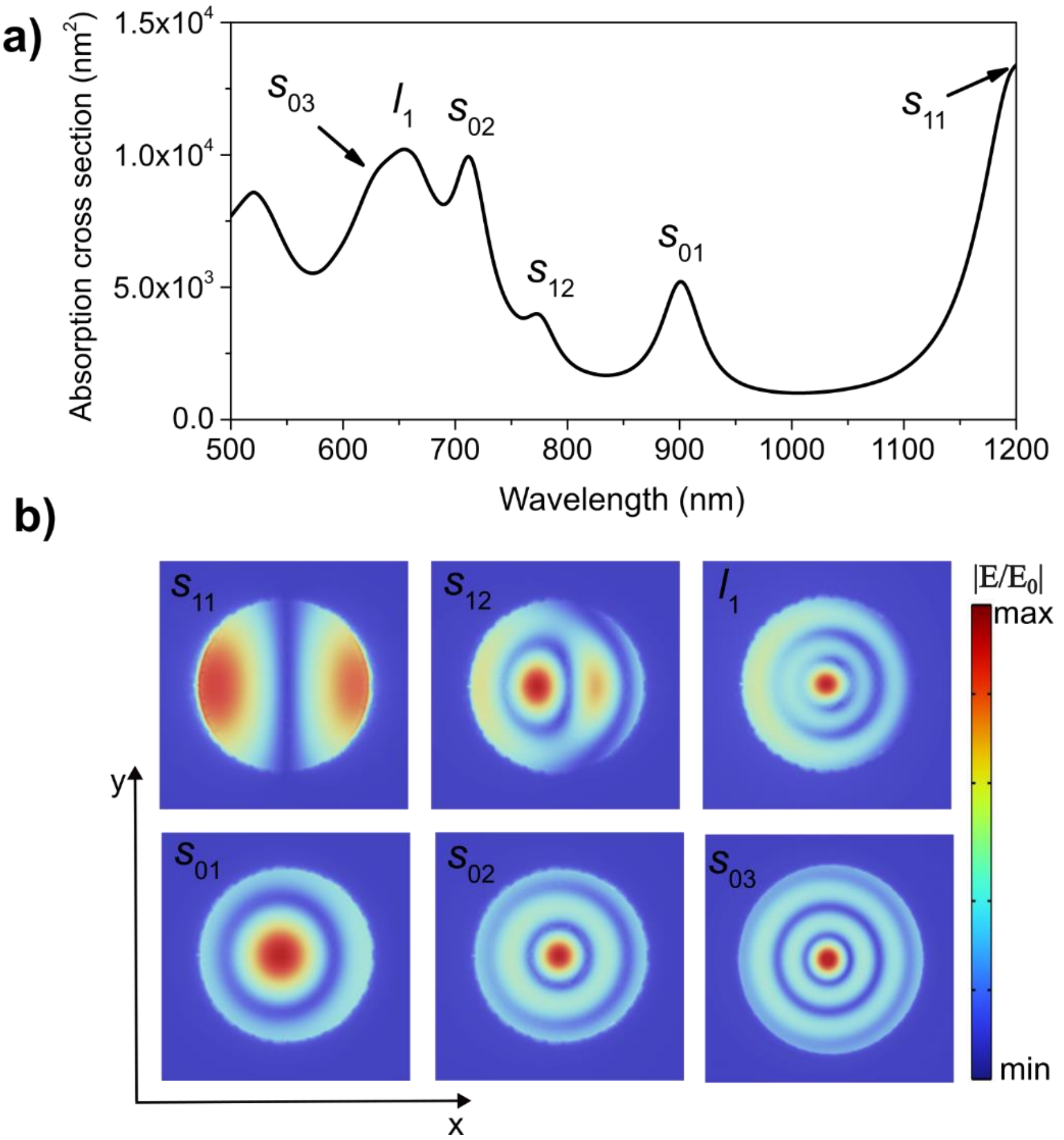

**Fig. S5 Near-field distribution of the TCP modes of the bare NPoM nanocavity.** (a) Absorption cross section spectra of the bare NPoM nanocavity (i.e., without picopatch) with gap thickness $d$ = 1.1 nm (other parameters are the same as in Fig. S4). (b) Corresponding near-field distributions in the horizontal x-y plane through the nanogap center (z = 0.55 nm, with axis defined in Fig. 1(c) in the main text) for LAP mode ($l_1$) and TCP $s_{mn}$ modes of different (m, n) orders. The results in (b) are obtained for wavelengths of $\lambda$ = 630 nm (mode $s_{03}$), $\lambda$ = 655 nm (mode $l_1$), $\lambda$ = 710 nm (mode $s_{02}$), $\lambda$ = 775 nm (mode $s_{12}$), $\lambda$ = 900 nm (mode $s_{01}$), and $\lambda$ = 1200 nm (mode $s_{11}$). We plot the enhancement of the field amplitude $|E/E_0|$ and the mode plotted in each panel is indicated by the labels.

The assignment of the NPoM eigenmodes can be further verified by plotting their near-field distributions for gap thickness $d$ = 1.1 nm. The E-field enhancement distributions ($|E/E_0|$) in the horizontal x-y plane at the middle of the gap are plotted in Fig. S5(b) for different absorption peaks, that is, at the wavelengths of the maxima of the absorption cross section shown in Fig. 5(a) (in the case of $s_{03}$, the position is estimated from the saddle point). The maxima in the cross section and the field plots are labelled $s_{03}$, $l_1$, $s_{02}$, $s_{12}$, $s_{01}$ and $s_{11}$ according to the discussion above. The field enhancement of the $s_{mn}$ modes follows to a good approximation the equation $|J_m(a'_{mn}\,\rho/r_{facet})\cos(m\varphi)|$ as discussed in the main text, confirming the modal assignment (notice that for these modes $|E_z/E_0|\approx|E/E_0|$). The match with the analytical equation of the fields is not perfect in large part because the simulated fields are determined not only by the mode that is resonantly excited, but also by other contributions, which explains the asymmetry of some of the calculated field distributions. We further note briefly that the near-field enhancement associated with the $s_{11}$ mode is close to zero at the gap center (Fig. S5(b)), which explains why there is no corresponding peak in the field enhancement spectra in Fig. S4(b). Last, the LAP mode $l_1$ presents a similar near-field distribution in the x-y plane to the $s_{03}$ mode, which is due to a relatively small detuning between these two modes (Fig. S5(a)), so that the $s_{03}$ mode contributes strongly to the cavity fields at the resonant wavelength of the $l_1$ mode.

## S4. Dependence on $r_{slot}$ of scattering spectra of NPoM resonator with picopatch

Figure 4(a) in the main text shows the evolution of the absorption cross section spectra and near-field enhancement spectra as we change the radius of the slot situated at the middle of the picopatch, for the configuration sketched in Fig. 1(c) in the main text. For completeness, Fig. S6 shows the corresponding results for the scattering cross section.

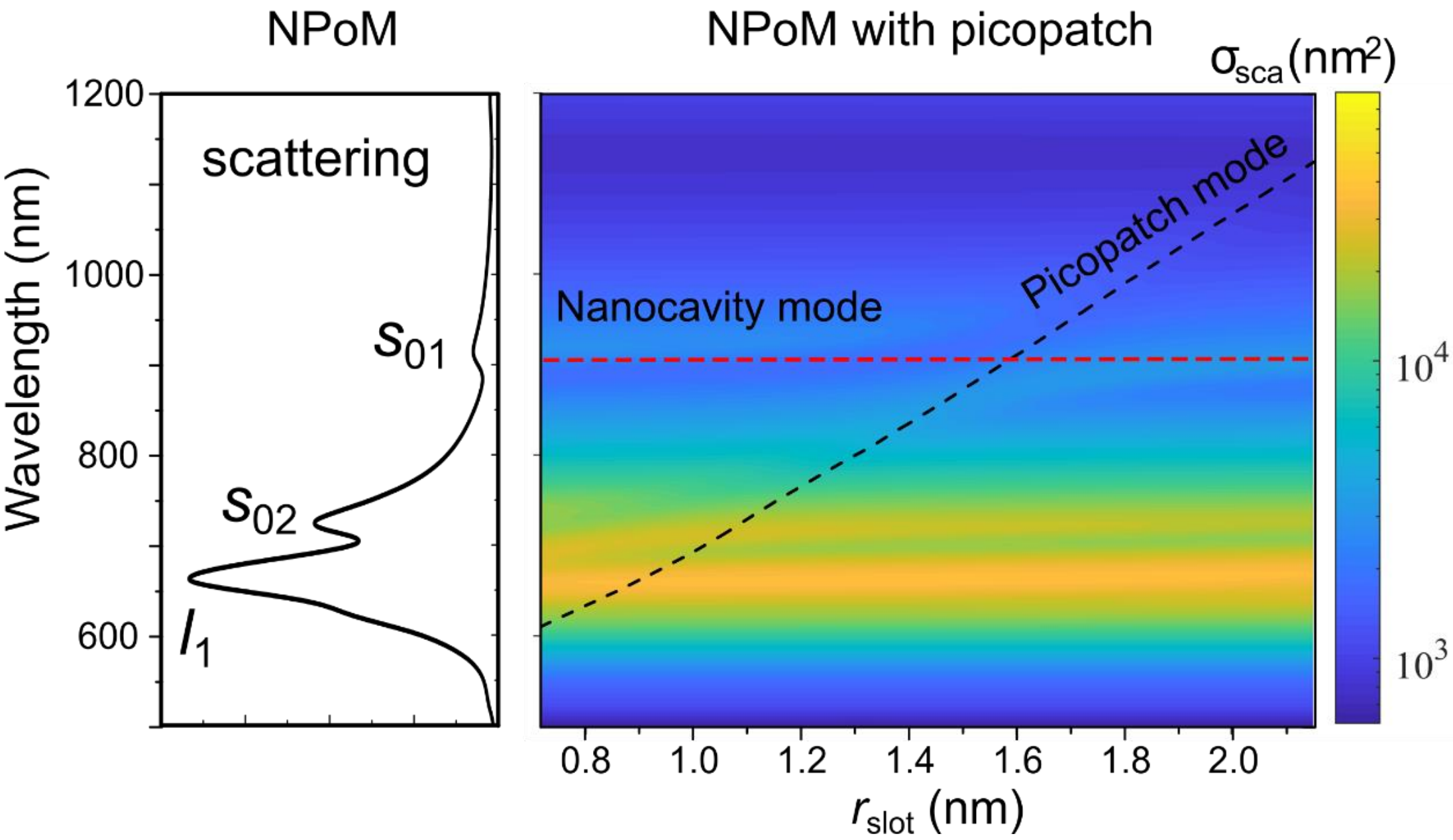

**Fig. S6 Scattering cross section of the NPoM resonator with picopatch.** Color maps of the scattering cross sections (in nm$^2$) of the NPoM containing a picopatch as a function of the wavelength $\lambda$ and the slot radius $r_{slot}$. The system is the same as in Fig. 4 of the main text. The spectra to the left of the color map corresponds to the results for the bare NPoM nanocavity, consisting of the same system after removing the picopatch. The horizontal dashed line in the color map indicates the position of the bare NPoM nanocavity eigenmode ($s_{01}$). The theoretical wavelength of the $s'_{01}$ picopatch mode introduced by the picopatch is shown by the diagonal dashed line, as obtained from Eqs. (2) and (4) in the main text that do not include coupling between modes.

## S5. Near-field distributions of higher-order picopatch TCP modes

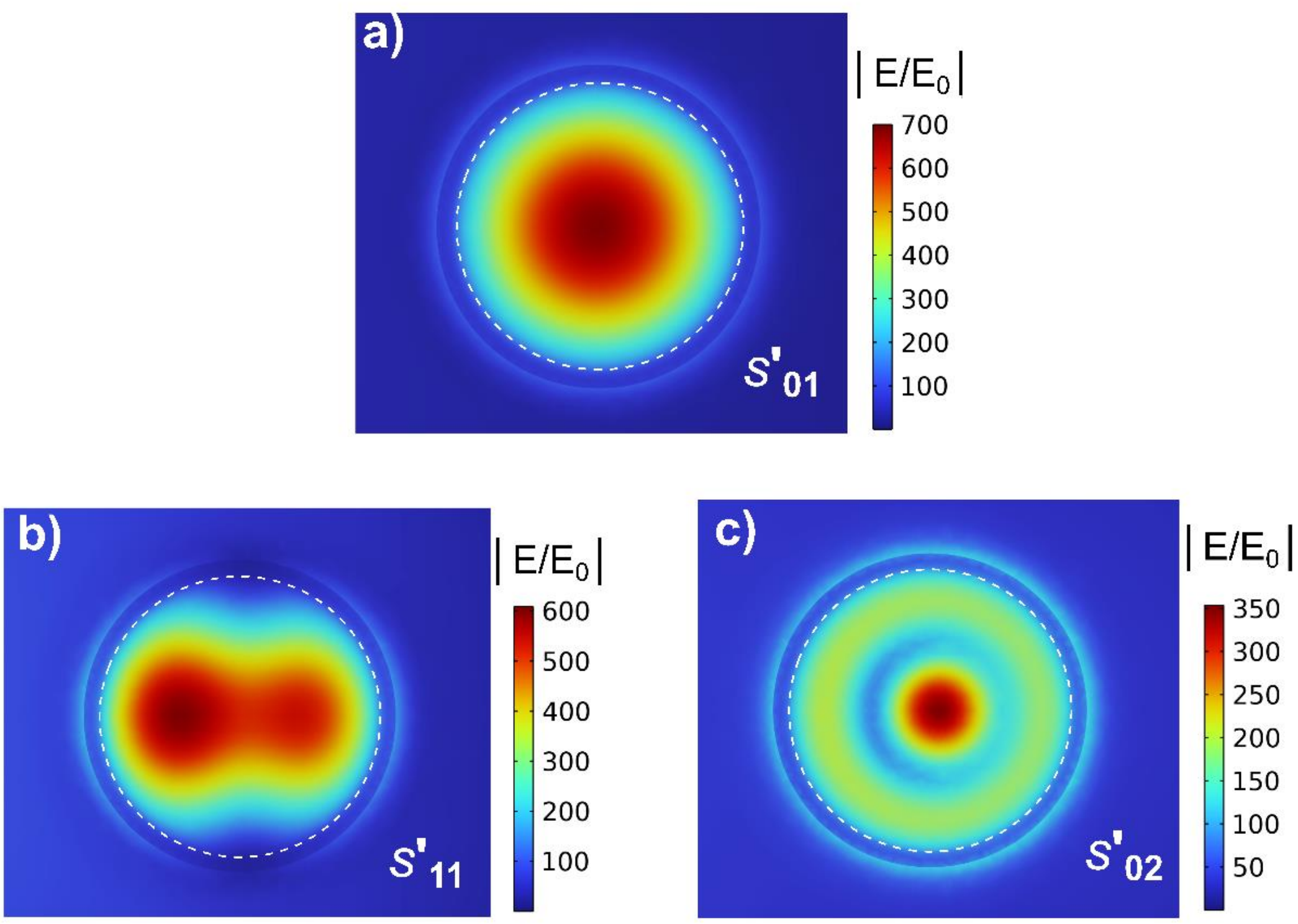

**Fig. S7 Near-field distributions of higher-order picopatch TCP modes.** We consider the NPoM resonator with picopatch, for the same geometry as in Figs. 5(c, d) in the main text (slot radius $r_{slot}$ = 2.153 nm). The enhancement of the near-field amplitude ($|E/E_0|$) is calculated in the horizontal x-y plane across the middle of the vacuum slot of the picopatch (top view, z = 0.1 nm, axis in Fig. 1(c) in the main text). The white dashed line corresponds to the edge of the slot. The fields are calculated at a wavelength (a) $\lambda$ = 1080 nm (marked with 'III' in Fig. 4(b) in the main text, and corresponding to the wavelength of the absorption peak), (b) $\lambda$ = 746 nm and (c) $\lambda$ = 616 nm. The wavelengths $\lambda$ = 746 nm and 616 nm correspond to the values predicted for the $s'_{11}$ and $s'_{02}$ modes, respectively, by Eq. (2) and Eq. (4) in the main text.

The analysis of the picopatch modes in the main text focuses on the $s'_{01}$ mode. However, the picopatch supports other higher-order modes. To illustrate them, we fix $r_{slot}$ = 2.153 nm and plot in Fig. S7 the field distribution in a horizontal x-y plane going through the center of the slot of the picopatch. The wavelength used in these calculations corresponds to the wavelength of the absorption peak for the $s'_{01}$ mode (Fig. S7(a)), and, for the $s'_{11}$ (Fig. S7(b)) and $s'_{02}$ (Fig. S7(c)) modes, to the resonant wavelength predicted by the theoretical model (Eq. (2), corresponding to Eq. (S1.7), and Eq. (4) in the main text, which neglect coupling). The field distribution for the $s'_{02}$ and $s'_{11}$ modes of the picopatch resembles the nanocavity modes of the same order in Fig. S5(b). The main difference between them is that the nanocavity modes present a maximum at the cavity edges, while the picopatch mode fields are small near the edge of the slot (indicated by white dashed line in Fig. S7). This difference is attributed to the change in boundary conditions (i.e., metallic wall at the end of the picopatch slot, but air at the edges of the nanocavity), which also explains the difference between the corresponding analytical equations of the resonant plasmonic wavelengths (Eqs. (3) and (4) in the main text).

## S6. Calculation of Purcell factor and extraction of effective volume

The Purcell factor $F_p$ represents the enhancement of the spontaneous decay rate $\gamma$ of an emitter placed in a region of permittivity $\varepsilon_d$ near a plasmonic or dielectric cavity, as compared to the corresponding value $\gamma_0$ when the emitter is placed in an infinite homogeneous medium of the same permittivity $\varepsilon_d$. We assume in this discussion that this medium is vacuum, which is consistent with the simulations in this work (except in section S9, where we consider the case of a different medium). The definition of $F_p$ is appropriate when the emitter couples weakly with the nanostructure. In the main text, we use $F_p$ as a second approach to obtain the effective volume of the modes, $V_{\mathrm{eff}}$, so that we can compare these values to those calculated by a direct integration of the fields (Eq. (6) in the main text). We discuss here the numerical simulation of $F_p$ and how to use it to extract $V_{\mathrm{eff}}$.

We consider a point dipole of transition dipole moment $\mu$ oriented along the z-axis (perpendicular to the gold surfaces at the gap, see sketch and axis in Fig. 1(c) in the main text). This dipole is situated at position $\vec{r}_m$ corresponding to the center of the picopatch slot (the same place at which the near-field enhancement spectra is often evaluated, as marked with a black cross in Fig. 1(c)). The position $\vec{r}_m$ and orientation are the same as when evaluating Eq. (6) to obtain $V_{\mathrm{eff}}$ from the field distribution. We assume that pure dephasing is negligible and that $\gamma_0$ is much smaller than the cavity losses.[12] The Purcell factor $F_p$ is then calculated from the z component of the electric fields induced by this dipole at the same $\vec{r}_m$ position of the dipole, $E_z(\vec{r}_m, \omega)$. More exactly, we have

$$F_p = \frac{\gamma}{\gamma_0} = \frac{6\pi c}{\omega} Im\{G_{zz}(\vec{r}_m, \vec{r}_m, \omega)\} = \frac{6\pi c^3 \varepsilon_0}{\omega^3 \mu} Im\{E_z(\vec{r}_m, \omega)\}, \qquad \text{(S6.1)}$$

where Im{} represents the imaginary part, $\omega$ the angular frequency of the dipole, $c$ the velocity of light in vacuum and $G_{zz}(\vec{r}_m, \vec{r}_m, \omega)$ the zz component of the self-interaction dyadic Green's function $\overleftrightarrow{G}(\vec{r}_m, \vec{r}_m, \omega)$. We obtain $E_z(\vec{r}_m, \omega)$ from the COMSOL simulations of the NPoM resonator with picopatch (Fig. 1(c) in the main text) and plot the Purcell factor obtained from Eq. (S6.1) in Fig. S8, as a function of the slot radius $r_{slot}$ and the wavelength of the dipole. Extremely large values are obtained, up to $F_p \approx 5 \times 10^8$, with a clear avoided crossing in the region near $r_{slot}$ ≈ 1.5 - 1.7 nm and $\lambda$ ≈ 900 nm.

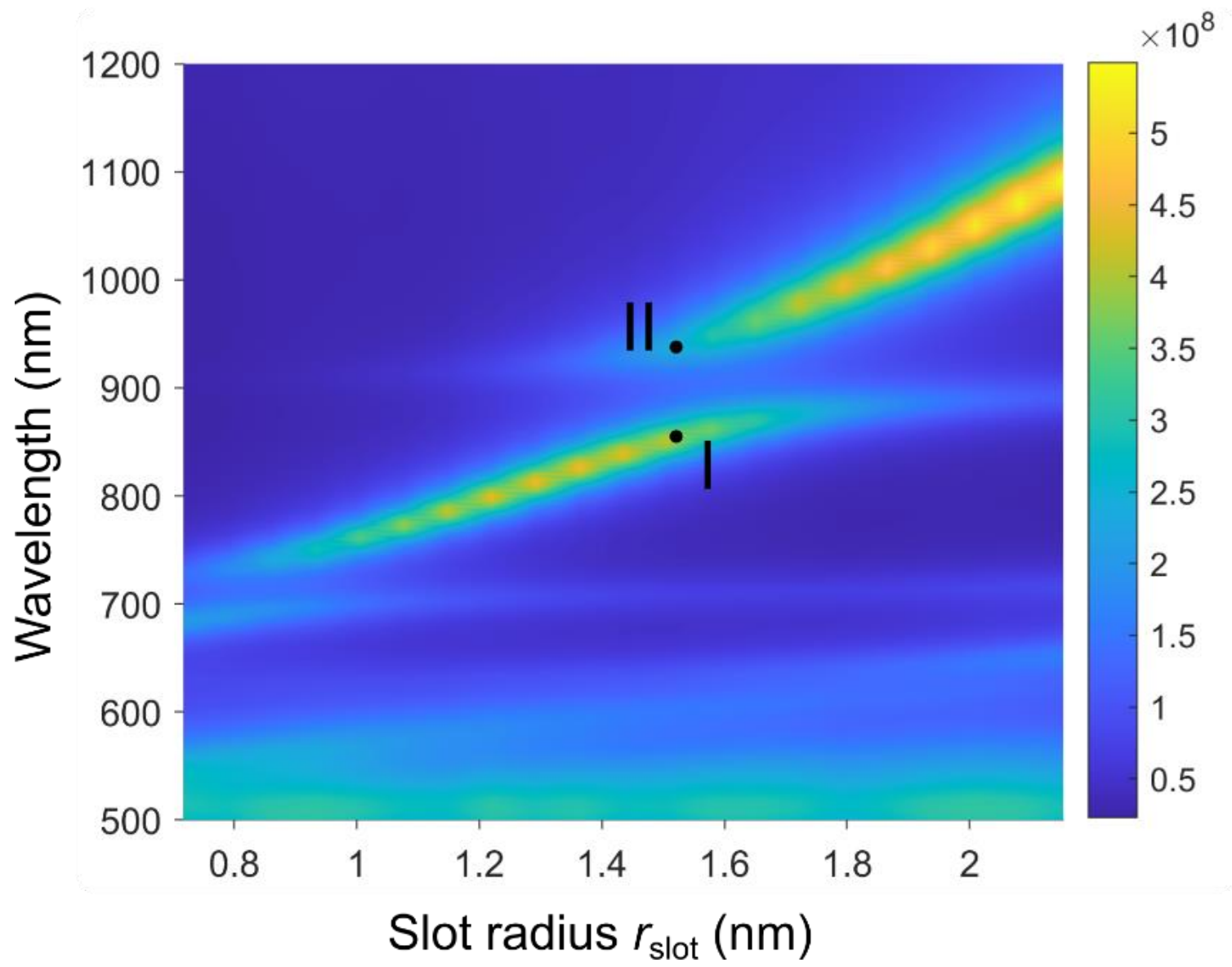

**Fig. S8 Purcell factor of the NPoM resonator with picopatch**. The Purcell factor obtained according to the procedure described in section S6 is plotted as a function of wavelength ($\lambda$) and slot radius ($r_{slot}$). The geometry is the same as in Fig. 6 in the main text, and the dipole used in the simulations is placed at the center of the picopatch slot (Fig. 1(c) in the main text). The dipole is oriented in the vertical z direction perpendicular to the gold surfaces forming the nanogap.

When a single, Lorentzian-like mode dominates the response, the Purcell factor at resonance follows [13,14]

$$F_p = \frac{3Q}{4\pi^2 \mathrm{V_{eff}}}(\lambda_r)^3, \tag{S6.2}$$

where $Q$ is the quality factor, $V_{eff}$ the effective mode volume, $\lambda_r$ the resonant wavelength (in vacuum) of the cavity mode, and we are again taking into account that $\varepsilon_d = 1$ at the position of the dipole. Eq. (S6.2) assumes that the emitter is placed at the position of the strongest fields, and polarized along the direction of these fields, which is confirmed to be true by the COMSOL calculations (to a very good approximation) for the calculations in this section and in the main text.

To obtain $F_p$ at resonance, we use real-valued Lorentz spectral lineshapes to fit the simulated Purcell factor spectra in Fig. S8 of the NPoM resonator with picopatch,

$$y = y_0 + \sum_{i=1}^{2} \frac{2A_i}{\pi} \frac{w_i}{4(x - x_{ci})^2 + {w_i}^2} \tag{S6.3}$$

with $x_{ci}$ the central position (angular frequency), $w_i$ the full width at half maximum (FWHM), $A_i$ the amplitude that determines the peak height, and $y_0$ the baseline value. We fit the spectra when plotted as a function of frequency.

We fit the spectra for each value of $r_{slot}$ and extract the values of $Q_i = \frac{x_{ci}}{w_i}$, $\lambda_{ri} = \frac{2\pi c}{x_{ci}}$ ($c$: speed of light in vacuum) and $F_p$ at resonance for the maxima that appear between $\lambda$ = 745 - 1200 nm for $r_{slot}$ = 1.220 - 2.153 nm. These maxima correspond to the hybrid modes discussed in the main text. Once all these values have been obtained, Eq. (S6.2) directly gives the effective volume $V_{eff}$ plotted in Fig. 6 in the main text (red lines). We present in Table S1 the values of $Q$, $\lambda_r$ and $F_p$ obtained from the fit for different slot radius, together with the resulting value of $V_{eff}$.

| $r_{slot}$ (nm) | Mode I | | | | Mode II | | | |
|---|---|---|---|---|---|---|---|---|
| | $\lambda_r\ (nm)$ | Fp (x$10^8$) | Q | $V_{eff}$ ($nm^3$) | $\lambda_r\ (nm)$ | Fp (x$10^8$) | Q | $V_{eff}$ ($nm^3$) |
| 1.220 | 800 | 4.42 | 22.6 | 1.99 | 915 | 0.79 | 22.33 | 16.39 |
| 1.292 | 815 | 4.47 | 22.53 | 2.07 | 920 | 1.03 | 19.63 | 11.33 |
| 1.363 | 825 | 4.53 | 22.48 | 2.12 | 925 | 1.37 | 18.70 | 8.24 |
| 1.435 | 840 | 4.43 | 22.62 | 2.30 | 930 | 1.80 | 17.67 | 5.99 |
| 1.506 | 850 | 4.14 | 22.91 | 2.59 | 935 | 2.35 | 17.30 | 4.58 |
| 1.578 | 860 | 3.66 | 23.41 | 3.09 | 950 | 3.00 | 17.07 | 3.70 |
| 1.650 | 870 | 3.15 | 24.01 | 3.82 | 960 | 3.61 | 17.05 | 3.18 |
| 1.722 | 875 | 2.63 | 24.75 | 4.79 | 980 | 4.13 | 17.13 | 2.97 |

| 1.794 | 880 | 2.18 | 25.98 | 6.17 | 995 | 4.53 | 17.05 | 2.82 |
|---|---|---|---|---|---|---|---|---|
| 1.865 | 885 | 1.85 | 27.94 | 7.96 | 1010 | 4.82 | 17.04 | 2.77 |
| 1.937 | 887 | 1.58 | 28.82 | 9.65 | 1030 | 5.06 | 16.37 | 2.69 |
| 2.009 | 890 | 1.39 | 28.61 | 11.02 | 1050 | 5.26 | 15.77 | 2.64 |
| 2.081 | 890 | 1.24 | 27.60 | 11.97 | 1070 | 5.39 | 15.53 | 2.68 |
| 2.153 | 895 | 1.10 | 25.73 | 12.69 | 1090 | 5.49 | 15.20 | 2.73 |

**Table S1. Resonant wavelength, quality factor and Purcell factor at the resonant wavelength, extracted from the fit of the simulations of the Purcell factor with Eq. (S6.3), as well as effective mode volume obtained from Eq. (S6.2).** The results are given for the two hybrid modes (labeled with I and II in Fig. S8 and Fig. 6 in the main text), for different slot radius.

## S7. Dependence on $r_{slot}$ of the scattering spectra of NPoM with non-flat picopatch

Figures 7(b, c) in the main text show the evolution of the absorption cross section and near-field enhancement spectra when we change the radius of the slot situated at the middle of the picopatch, where the shape of this picopatch is a semi-ellipsoid (as given by the sketch in Fig. 7(a) in the main text). For completeness, in Fig. S9, we present the corresponding results for the scattering cross section.

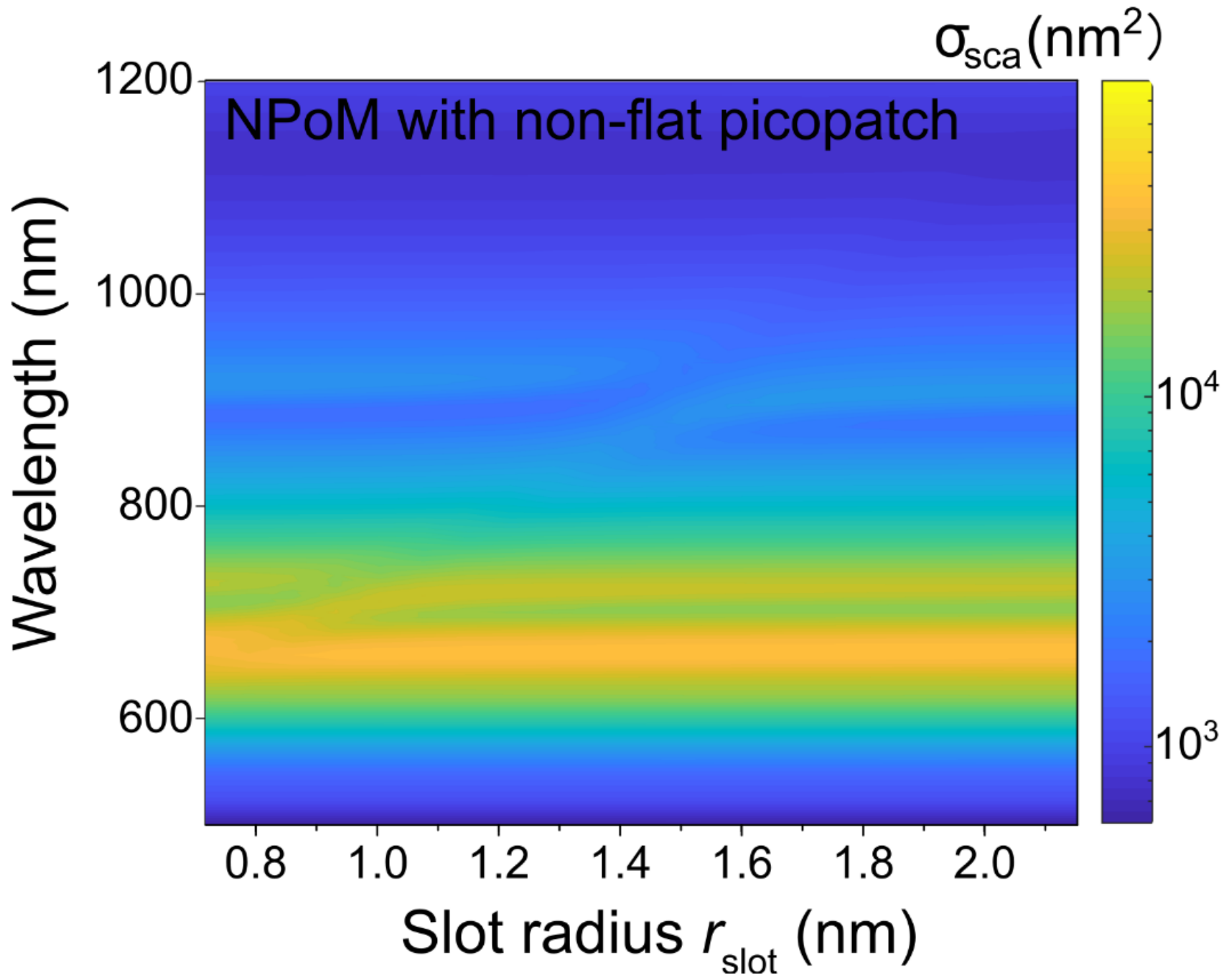

**Fig. S9 Color maps of the scattering cross section (in nm$^2$) of the NPoM resonator with a picopatch of a semi-elliptical shape (as sketched in Fig. 7(a) in the main text).** The color map is calculated as a function of the wavelength $\lambda$ and the picopatch slot radius $r_{slot}$.

**S8. Effect of increased absorption losses on the near-field enhancements in the middle of the nanogap**

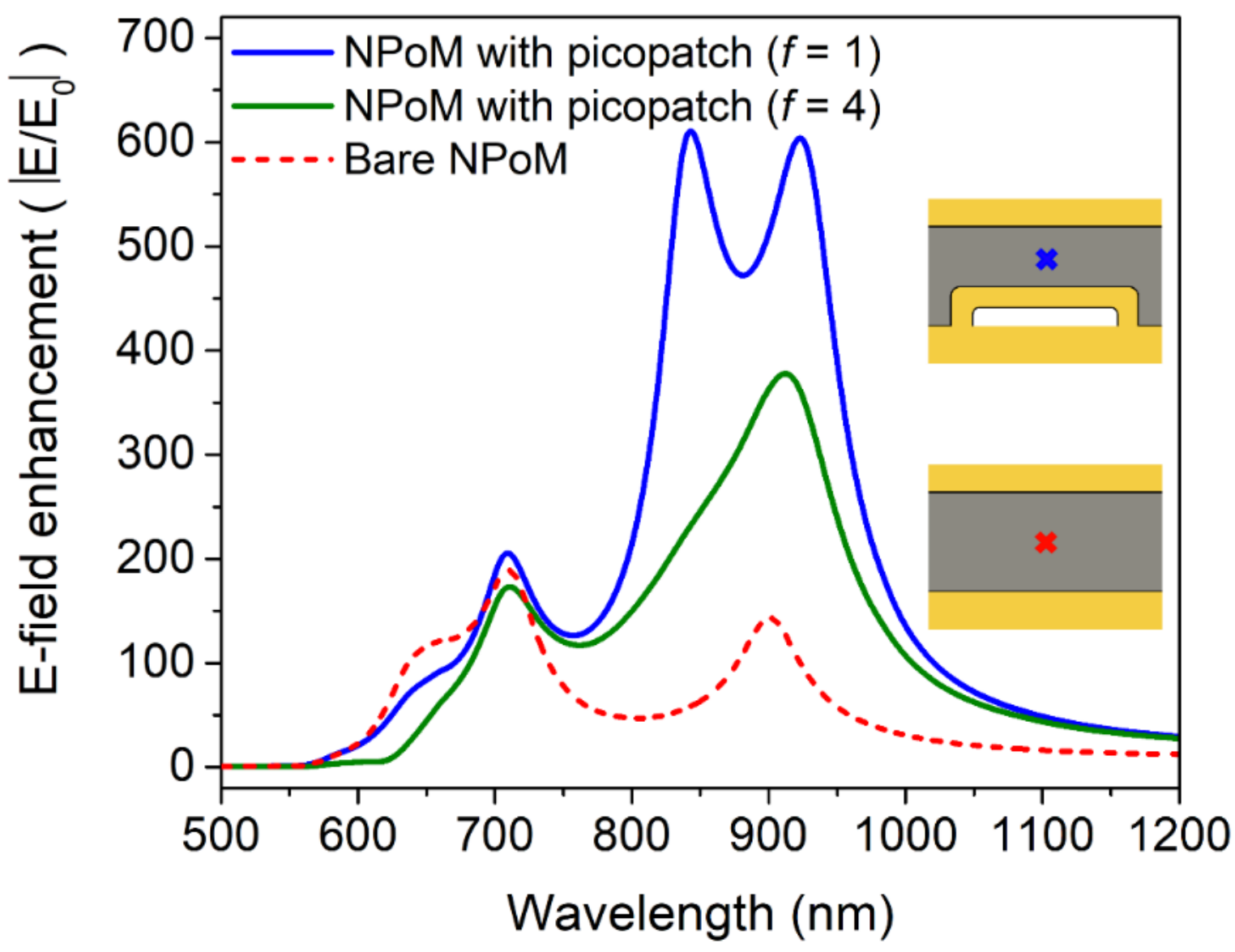

**Fig. S10 Effect of increased absorption losses on the near-field enhancements evaluated in the middle of the nanogap.** (Solid lines) Electric-field enhancement induced by the NPoM with a picopatch, evaluated in the center of the nanogap situated above the Au monolayer that forms the top of the picopatch (marked by blue cross in top inset, 0.7675 nm over the surface of the bottom substrate, or 0.3325 nm over the top of the picopatch monolayer). The results are obtained using the standard Au permittivity (blue solid line, corresponding to the blue line in Fig. 3(c) in the main text), and considering increased absorption losses in the gold permittivity (green solid line, $f = 4$ in Eq. (7) in the main text). (Red dashed line) Electric-field enhancement evaluated in the center of the nanogap for the bare NPoM without picopatch (marked by the red cross in the bottom inset, corresponding to a point at 0.55 nm over the surface of the bottom substrate, corresponding to red dashed line in Fig. 3(c) in the main text). The results are obtained for the following parameters: $r_{sphere}$ = 40 nm, $r_{facet}$ = 30 nm, $d$ = 1.1 nm, $\delta$ = 0.2 nm, $t$ = 0.235 nm, and $r_{slot}$ = 1.435 nm.

We have shown in the main text the effect of increasing the absorption losses in the slot region formed inside the picopatch, to consider possible phenomena not accounted for by the classical calculations. The losses were included by following the approach described in Eq. (7) and Figs. 7(d-f) in the main text. For reference, we consider in this section the effect of the losses on the E-field enhancement spectra evaluated at the position above the picopatch.

Figure S10 shows the full E-field enhancement spectra near the middle of the nanogap for the NPoM resonator with and without a picopatch of $r_{slot}$ = 1.435 nm radius (the exact position where the E-fields are evaluated is slightly different for both structures, as discussed in the caption). As we showed in the main text, for the experimental permittivity of gold, the presence of the picopatch leads to a strong enhancement of the maximum of the E-field spectra in the gap region over the picopatch ($|E/E_0| \approx 600$ at $\lambda = 845$ nm and at $\lambda = 925$ nm, blue solid line), compared to $|E/E_0| \approx 200$ at $\lambda = 710$ nm and $|E/E_0| \approx 150$ at $\lambda = 900$ nm in the case of the bare NPoM (red dashed line). Furthermore, if we increase the absorption losses in the picopatch region to consider possible effects not accounted for by the classical calculations (following the approach considered in Eq. (7) and Figs. 7(d-f) in the main text, for $f = 4$), the maximum of the E-field spectra is still very substantially enhanced (green solid line) compared with the results of the bare NPoM resonator. The results confirm that the picopatch could induce a strong near-field enhancement also in the nanogap region above the picopatch.

**S9. Purcell factor calculated at different positions**

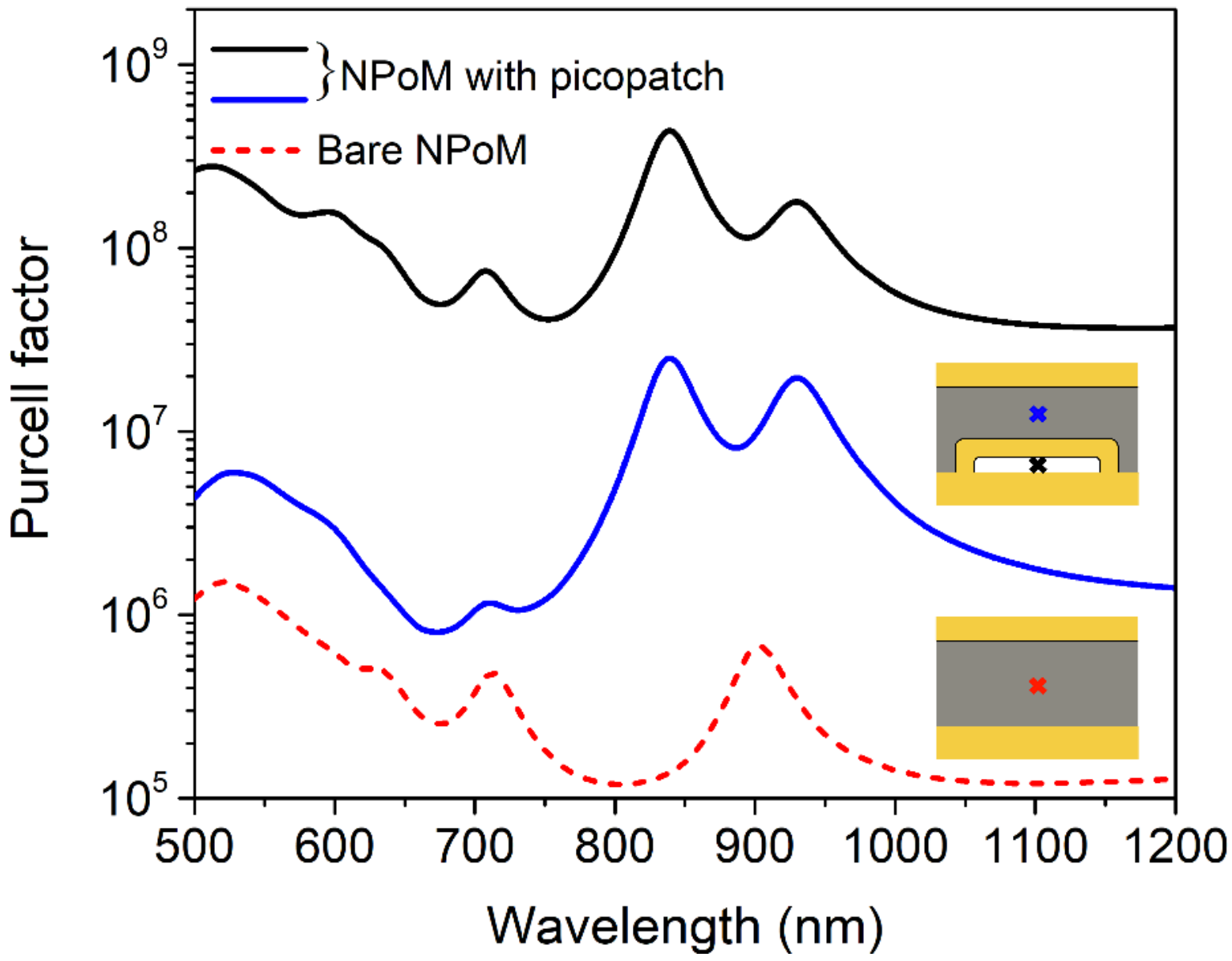

**Fig. S11 Purcell factor calculated at different positions.** (Solid lines) Purcell factor of an emitter situated at different positions in a NPoM with a picopatch of $r_{slot}$ = 1.435 nm radius. The emitter is situated at the center of the closed slot inside the picopatch (black solid line, corresponding to the $r_{slot}$ = 1.435 nm spectra in Fig. S8; position marked by black cross in top inset) and at the center of the nanogap situated above the Au monolayer that forms the top of picopatch (blue solid line; position marked by blue cross in top inset, 0.7675 nm over the surface of the bottom substrate or 0.3325 nm over the top of the picopatch monolayer). (Red dashed line) Purcell factor of an emitter placed at the center of the nanogap for the bare NPoM without picopatch (position marked by red cross in the bottom inset, 0.55 nm over the surface of the bottom substrate). The emitter is always polarized in the vertical z direction. The results are obtained for the following parameters: $r_{sphere}$ = 40 nm, $r_{facet}$ = 30 nm, $d$ = 1.1 nm, $\delta$ = 0.2 nm, $t$ = 0.235 nm, and $r_{slot}$ = 1.435 nm.

The Purcell factor of an emitter situated at different positions, calculated using the same procedure given in section S6 for $r_{slot}$ = 1.435 nm, is shown in Fig. S11. The Purcell factor at the middle of the nanogap (see caption for discussion of the exact positions) exhibits a maximum value of ≈ $2.5\times10^7$ (blue solid line), about one order of magnitude smaller than the corresponding Purcell factor of ≈ $4.5\times10^8$ for the same structure but evaluated in the center of the vacuum slot inside the picopatch (black solid line). The value of $2.5\times10^7$ remains ≈ 40 times larger than the Purcell factor of ≈ $6.7\times10^5$ of the bare NPoM without picopatch, evaluated in the center of the nanogap (red dashed line). The results indicate that in the presence of the picopatch the Purcell factor remains very large even at the nanogap region.

## S10. Calculation of the effective mode volume using quasinormal modes with imaginary frequencies

In the main text, to determine the effective mode volume ($V_{\text{eff}}$) through Eq. (6), we evaluate the fields at the real-valued resonant frequencies that correspond to the maxima of the absorption spectra (after artificially reducing the material losses by a factor 10 to narrow the resonant peaks). However, within the Quasi-Normal Modes (QNM) formalism, a more rigorous calculation of $V_{\text{eff}}$ with Eq. (6) requires computing the natural modes of the open and lossy system, which are characterized by complex-valued eigenfrequencies.[15-17] Here, we compare the two approaches to validate the strategy used in the main text.

To calculate the QNMs at complex frequencies, we describe the gold permittivity using a Drude-Lorentz model

$$\varepsilon_{Au,\text{Drude}}(\omega) = \varepsilon_{\infty}\left(1-\frac{\omega_{p,1}^2}{\omega^2+i\gamma_1\omega}-\frac{\omega_{p,2}^2}{\omega^2-\omega_{0,2}^2+i\gamma_2\omega}\right), \quad \text{(S10.1)}$$

where $\omega_{p,1}$ is the plasma frequency, $\omega_{p,2}$ sets the oscillator strengths of the Lorentz resonance, $\gamma_1$ and $\gamma_2$ are the damping rates, $\omega_{0,2}$ is the Lorentz resonance frequency, and $\varepsilon_{\infty}$ is the high-frequency permittivity. We obtain the values of these parameters by fitting this model to the experimental permittivity data[18] in the spectral range of interest (550-1100 nm), which gives

$$\varepsilon_{\infty} = 8.84 \quad \omega_{p,1} = 4.52\cdot 10^{15}\,\frac{rad}{s} \quad \omega_{p,2} = 0.42\cdot 10^{15}\,\frac{rad}{s} \quad \gamma_1 = 0.11\cdot 10^{15}\,\frac{rad}{s}$$

$$\gamma_2 = 0.48\cdot 10^{15}\,\frac{rad}{s} \quad \omega_{0,2} = 3.19\cdot 10^{15}\,\frac{rad}{s}\ . \quad \text{(S10.2)}$$

To assess the effect of using Eq. (S10.1) to fit the permittivity, we compare in Fig. S12 the field-enhancement spectra obtained for the experimental permittivity[18] $\varepsilon_{Au,\text{exp}}$ (solid black line) and the fitted Drude-Lorentz permittivity $\varepsilon_{Au,\text{Drude}}$ (dotted black line) for the geometry described in Fig. 1(c) of the main text with a slot radius $r_{\text{slot}}$ = 1.435 nm (note that $\varepsilon_{Au,\mathrm{exp}}$ has been denoted $\varepsilon_{Au}$ in the rest of the paper). The spectra obtained within these two approaches are not identical because $\varepsilon_{Au,\mathrm{Drude}}$ is not a perfect fit to $\varepsilon_{Au,\mathrm{exp}}$, but the differences are relatively small, which validates the use of $\varepsilon_{Au,\mathrm{Drude}}$ in this section to analyze the modes within the QNM formalism.

We compute the resulting QNMs of this NPoM resonator with picopatch using the MAN: Eig software of the freeware MAN (Modal Analysis of Resonators) under the COMSOL Multiphysics environment.[19] This method provides the complex-valued natural eigenfrequencies and eigenmodes of the system. For the geometry shown in Fig. 1(c) of the main text with $r_{\mathrm{slot}}$ = 1.435 nm, the complex-valued angular eigenfrequencies of the I and II modes are $\widetilde{\omega_{\mathrm{I}}} = (2.22 - \mathrm{i}\ 0.055) \cdot 10^{15}\,\frac{\mathrm{rad}}{\mathrm{s}}$ and $\widetilde{\omega_{\mathrm{II}}} = (2.01 - \mathrm{i}\ 0.052) \cdot 10^{15}\,\frac{\mathrm{rad}}{\mathrm{s}}$.

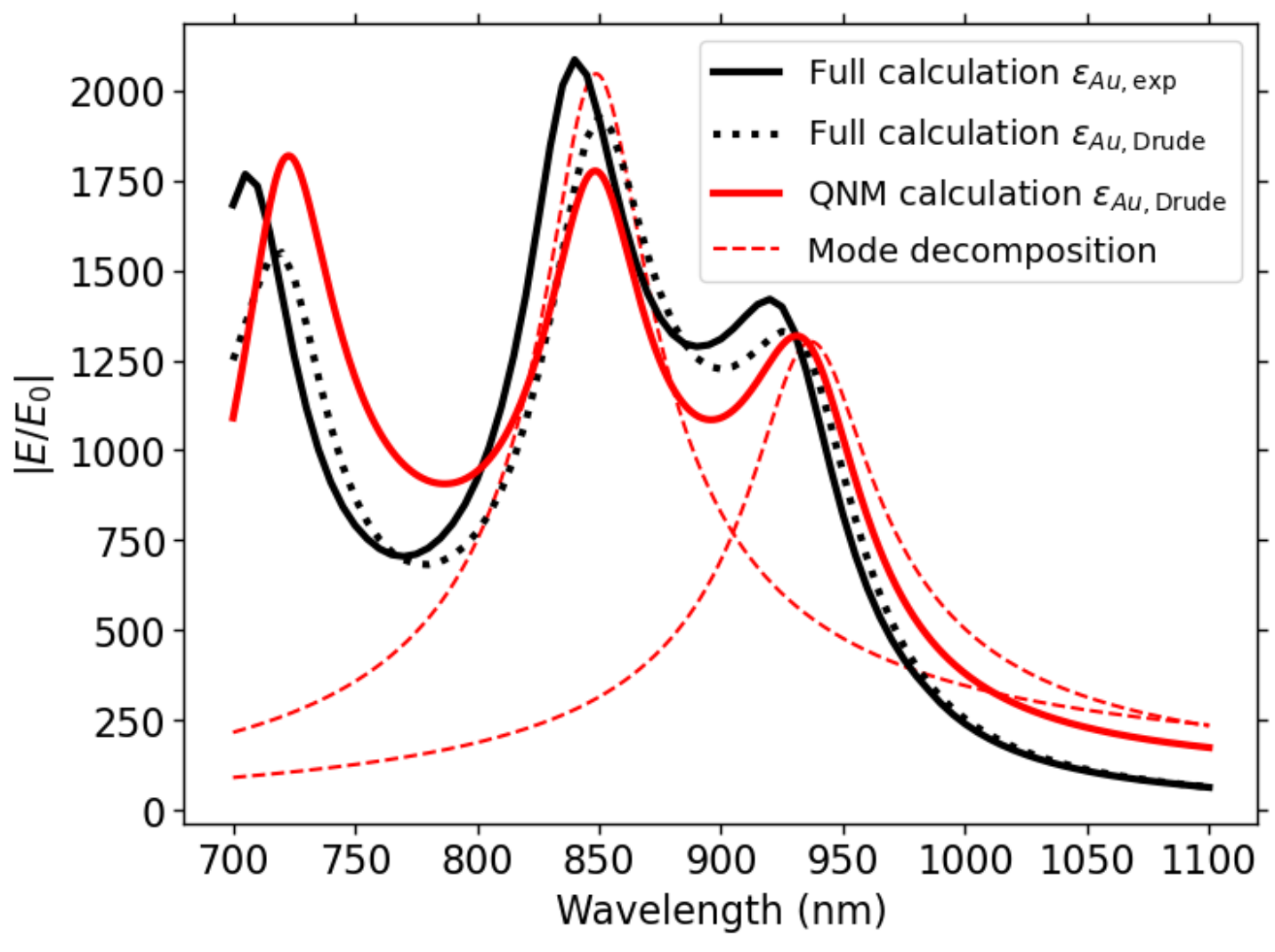

Fig. S12 **Comparison of the field-enhancement spectra obtained from the full-wave simulations and from the quasinormal modes with complex eigenfrequencies.** Near-field spectra obtained from the full-wave COMSOL calculations using the experimental permittivity of gold (solid black line, same as Fig. 3(c) of the main text); from the full-wave COMSOL calculations using the fitted Drude-Lorentz model for the gold permittivity (dotted black line); and from the QNM formalism including only the three modes that contribute most strongly in the 700-1100nm spectral range (red solid line). The red dashed lines correspond to the individual contributions to the near-field enhancement of the modes I and II extracted from the QNM expansion. The structure corresponds to that shown in Fig. 1(c) with $r_{\mathrm{slot}}$ = 1.435 nm, and the field enhancement is obtained at the center of the picopatch slot (black cross in Fig. 1(c)). The value of other parameters is: $r_{\mathrm{sphere}}$ = 40 nm, $r_{\mathrm{facet}}$ = 30 nm, $d$ = 1.1 nm, $\delta$ = 0.2 nm and $t$ = 0.235 nm.

We verify next that these QNM results enable the reconstruction of the optical response of the picopatch under the same plane-wave illumination used in the main text. Within this QNM framework, the system response is approximated as a linear superposition of its eigenmodes. Figure S12 (solid red line) shows the field-enhancement spectra calculated at the center of the picopatch and reconstructed from the QNM formalism by including the three modes that contribute most strongly to the optical response. This mode reconstruction shows good overall agreement with the full-wave simulations that use the same permittivity (compare red solid and black dashed lines). The small discrepancies occur because we only include the three dominant modes (among them the hybrid modes I and II) in the QNM reconstruction, neglecting others.

Importantly, the QNM formalism makes it possible to isolate the contribution of modes I and II (red dashed lines). Applying Eq. (6) to the fields of these quasinormal modes with complex eigenfrequencies and taking the real part of the results, we obtain $V_{\mathrm{eff}}$ = 2.32 nm$^3$ and $V_{\mathrm{eff}}$ = 9.32 nm$^3$ for modes I and II, respectively. These values are remarkably close to those calculated in the main text using the real-valued eigenfrequencies obtained from the maxima of the absorption spectra ($V_{\mathrm{eff}}$ = 2.27 nm$^3$ and $V_{\mathrm{eff}}$ = 8.29 nm$^3$, respectively). This very satisfactory agreement strongly supports our premise that, for the system considered in this work, using real-valued resonance frequencies provides a fairly accurate estimation of the effective mode volume.